\documentclass[journal=jpcafh,manuscript=article]{achemso}
\usepackage{amsmath}
\usepackage{amssymb}
\usepackage[version=4]{mhchem}
\usepackage{xcolor}
\usepackage{hyperref}
\usepackage{mathtools,xparse}

\newcommand{\comm}[2]{ \left[ #1 , #2 \right ]}
\newcommand{\ket}[1]{ \left| #1 \right \rangle}
\newcommand{\bra}[1]{ \left\langle #1 \right |}
\DeclarePairedDelimiter{\norm}{\lVert}{\rVert}

\author{Peter Reinholdt}
\affiliation[SDU]{Department of Physics, Chemistry and Pharmacy, University of Southern Denmark, Campusvej~55, DK--5230 Odense M, Denmark}
\email{reinholdt@sdu.dk}
\author{Erik Kjellgren}
\affiliation[SDU]{Department of Physics, Chemistry and Pharmacy, University of Southern Denmark, Campusvej~55, DK--5230 Odense M, Denmark}
\author{Karl Michael Ziems}
\affiliation[DTU]{Department of Chemistry, Technical University of Denmark, Kemitorvet Building 207, DK-2800 Kongens Lyngby, Denmark.}

\author{Sonia Coriani}
\affiliation[DTU]{Department of Chemistry, Technical University of Denmark, Kemitorvet Building 207, DK-2800 Kongens Lyngby, Denmark.}
\author{Stephan P. A. Sauer}
\affiliation[KU]{Department of Chemistry, University of Copenhagen, DK-2100 Copenhagen \O.}
\author{Jacob Kongsted}
\affiliation[SDU]{Department of Physics, Chemistry and Pharmacy, University of Southern Denmark, Campusvej~55, DK--5230 Odense M, Denmark}

\title{Self-consistent Quantum Linear Response with a Polarizable Embedding environment}

\begin{document}


\begin{abstract}
Quantum computing presents a promising avenue for solving complex problems, particularly in quantum chemistry, where it could accelerate the computation of molecular properties and excited states.
This work focuses on hybrid quantum-classical algorithms for near-term quantum devices, combining the quantum linear response (qLR) method with a polarizable embedding (PE) environment. We employ the self-consistent operator manifold of quantum linear response (q-sc-LR) on top of a unitary coupled cluster (UCC) wave function in combination with a Davidson solver. The latter removes the need to construct the entire electronic Hessian, improving computational efficiency when going towards larger molecules.
We introduce a new superposition-state-based technique to compute Hessian-vector products and show that this approach is more resilient towards noise than our earlier gradient-based approach.
We demonstrate the performance of the PE-UCCSD model on systems such as butadiene and para-nitroaniline in water and find that PE-UCCSD delivers comparable accuracy to classical PE-CCSD methods on such simple closed-shell systems. We also explore the challenges posed by hardware noise and propose simple error correction techniques to maintain accurate results on noisy quantum computers.
\end{abstract}

\section{Introduction}
Quantum computing has the potential to solve certain computational problems significantly faster than classical computer architectures by leveraging properties such as quantum superposition and entanglement.  
One promising area where quantum computers might be useful is within quantum chemistry, particularly in solving the electronic Schr{\"o}dinger equation.
Future fault-tolerant quantum computers could, for example, be used to obtain ground state energies with quantum phase estimation\cite{kitaev1995quantum,aspuru2005simulated}. However, near-term quantum computers are rather limited in terms of qubit counts and execution speed. More importantly, hardware noise prevents any meaningful execution of deep quantum circuits required for the most promising quantum algorithms\cite{nash2020quantum,lau2022nisq}.
Thus, hybrid algorithms that leverage both quantum and classical computing architectures, allowing for shallower quantum circuits, are more realistic options for practical quantum computing applications in the near term. 

Much attention has been paid to solving the ground-state electronic Schr{\"o}dinger equation, in particular with the variational quantum eigensolver (VQE) algorithm\cite{peruzzo2014variational,mcclean2016theory}. Though less focus has been given to the treatment of general molecular properties and excited states, there is an increasing amount of proposed approaches for accessing excited states. These include quantum equation of motion or linear response approaches (qEOM/qLR\cite{ollitrault2020quantum,asthana2023quantum,kumar2023quantum,ziems2024options,reinholdt2024subspace,jensen2024quantum}), quantum subspace expansion (QSE\cite{mcclean2017hybrid,colless2018computation,mcclean2020decoding}), multi-state or state-averaged VQE methods\cite{nakanishi2019subspace,parrish2019quantum,yalouz2021state,fitzpatrick2024self,grimsley2024challenging}, and penalty-based methods such as variational quantum deflation (VQD)\cite{higgott2019variational}, among others.

Most chemistry occurs not in isolation but rather in solution or, more generally, within some larger chemical environment. 
Although methods leveraging quantum computers may eventually be able to treat very large molecular systems with a full quantum-mechanical description, describing the complete chemical environment
will most likely remain out of reach for many years to come, even with significant developments in both quantum hardware and algorithms. Fortunately, many chemical processes and events can reasonably be described as local processes, where the effect of the environment can accurately be included at a lower (and cheaper) level of theory.
In conventional quantum chemistry, this has long been recognized, leading to a wealth of environment models with various levels of complexity, expense, and accuracy. 
These models are characterized by combining quantum mechanics (QM) for the reactive part of a system with a simplified description of the surrounding environment, often based on molecular mechanics (MM) through classical force fields, which are computationally less expensive.

Among the simpler yet useful models are the class of continuum solvation models\cite{tomasi2005quantum}. In such models, the solvent is treated as an infinite, structureless continuum in which the solute is embedded in a cavity. The degrees of polarization response of this medium are determined by a dielectric constant; specifically, the dielectric response is modeled by an induced charge density on the surface of the cavity.
Continuum models work well in describing many simple solute--solvent systems and are particularly useful due to their computational efficiency, ease of use, and wide availability. On the other hand, they may be inadequate for describing, e.g., structured environments and directional molecular interactions, such as hydrogen bonds or interactions in biological systems, where specific solvent molecules play crucial roles.

For such cases, atomistic models can be an efficient and accurate option. Among the simpler models in this category are those based on electrostatic embedding\cite{singh1986combined,field1990combined}. Here, the QM charge density is polarized by the permanent charge distribution of the environment, often represented in terms of point charges. More advanced models also allow for mutual polarization to occur between the environment and the QM density. For example, the induced charge density of the environment is represented by induced dipoles\cite{olsen2010excited,olsen2011molecular,thompson1996qm,gordon2007effective,loco2016qm} or fluctuating charges\cite{Lipparini2011,lipparini2012analytical,lipparini2012linear}.
Beyond polarizable environment models are ``proper'' QM-QM embeddings, with prominent examples being frozen density embedding\cite{wesolowski1997kohn}, projector-based embedding\cite{manby2012simple}, and multi-level methods\cite{myhre2014multi}. 
Such models have the potential to be more accurate than simpler atomistic models but are typically also more complex and computationally costly.

In the context of quantum computing, many of these models have already been adapted to quantum computing frameworks. 
For example, the polarizable continuum model has been combined with VQE \cite{castaldo2022quantum}, which has been applied to study tautomeric state prediction in solution\cite{shee2023quantum}.
Beyond continuum solvation models, atomistic quantum mechanics/molecular mechanics (QM/MM) models have been explored with an electrostatic (point-charge) embedding for double-factorized Hamiltonians \cite{hohenstein2023efficient} and within a polarizable embedding with the PE-VQE-SCF\cite{kjellgren2024variational} framework. The latter was also applied to model electric field gradients for quadrupole coupling constants for ice VIII and ice IX \cite{nagy2024electric}.
More advanced quantum embedding models have also been explored \cite{rossmannek2023quantum,weisburn2024multiscale,ettenhuber2024calculating,battaglia2024general}. 

So far, the above-described solvation and embedding models within quantum computing have mostly been concerned with the treatment of 
 ground-state energies, occasionally including ground-state properties.
However, going beyond the ground state with a description that can also include excited states and molecular response theory is of utmost importance as it enables the modeling of, for example, absorption spectra and photochemical processes.
In this work, we focus on the so-called quantum linear response methods\cite{kumar2023quantum}, in particular, using the self-consistent operator manifold proposed by \citeauthor{asthana2023quantum}\cite{asthana2023quantum}, and combine it for the first time with a polarizable embedding environment description.
We rely on our recent implementation of a Davidson solver for q-sc-LR UCCSD\cite{reinholdt2024subspace}, which requires only the computation of hessian-vector products, which allows us to avoid the computationally intensive task of constructing the full Hessian matrix, improving efficiency when scaling towards larger systems.

\section{Theory}
\subsection{Unitary Coupled Cluster}
We first consider unitary coupled cluster (UCC) wave functions for molecules in isolation, which are parametrized as

\begin{eqnarray}
    \ket{\Psi_\textrm{UCC}} =  \prod_I e^{\hat{T}_{I}(\theta_I)-\hat{T}_{I}^{\dagger}(\theta_I)}\ket{0} = U(\boldsymbol{\theta}) \ket{0}
\end{eqnarray}
with the UCC wave function taking a factorized form to be representable on a quantum circuit, and $\ket{0}$ is the reference state (taken to be the Hartree-Fock state). 
We use spin-adapted operators \cite{Paldus1977,Piecuch1989,Packer1996}, which ensures spin conservation and reduces the dimension of the following quantum linear response generalized eigenvalue equations (\emph{vide infra}). 
The index $I$ runs over the selected set of excitation operators, which in this work is truncated to singles and doubles as
\begin{align}
    \ \hat{T}_{i}^{a}\left(\theta_i^a\right) &= \theta_i^a \frac{1}{\sqrt{2}}\hat{E}_{ai} \label{eq:wf_par_1}\\
    \ \hat{T}_{ij}^{ab}\left(\theta_{ij}^{ab}\right) &= \theta_{ij}^{ab} \frac{1}{2\sqrt{\left(1+\delta_{ab}\right)\left(1+\delta_{ij}\right)}}\left(\hat{E}_{ai}\hat{E}_{bj} + \hat{E}_{aj}\hat{E}_{bi}\right)\\
    \ \hat{T}_{\phantom{'}ij}^{'ab}\left(\theta_{\phantom{'}ij}^{'ab}\right) &= \theta_{\phantom{'}ij}^{'ab} \frac{1}{2\sqrt{3}}\left(\hat{E}_{ai}\hat{E}_{bj} - \hat{E}_{aj}\hat{E}_{bi}\right).
    \label{eq:wf_par_2}
\end{align}
Here, the indices $i,j$ refer to occupied orbitals in the reference state, while $a,b$ refer to virtual orbitals, and $\hat{E}_{pq} = a^\dagger_{p\alpha} a_{q\alpha} + a^\dagger_{p\beta} a_{q\beta} $ is a singlet one-electron excitation operator.

The wave function parameters ($\boldsymbol{\theta}$) are obtained using a variational minimization with a quantum eigensolver (VQE) procedure\cite{peruzzo2014variational}, 
\begin{equation}
    E_\text{0} = \min_{\boldsymbol{\theta}}\left<\Psi_{\mathrm{UCC}}\left(\boldsymbol{\theta}\right)\left|\hat{H}\right|\Psi_{\mathrm{UCC}}\left(\boldsymbol{\theta}\right)\right>~.
\end{equation}
Thus, we iteratively optimize the parameters to minimize the energy, with each iteration involving the measurement of the expectation value of the fermionic Hamiltonian
\begin{equation}
    \hat{H} = \sum_{pq}h_{pq}\hat{E}_{pq} + \frac{1}{2}\sum_{pqrs}g_{pqrs}\hat{e}_{pqrs}~,
\end{equation}
on the quantum computer.
Here, $h_{pq}$ and $g_{pqrs}$ are the one- and two-electron integrals and $\hat{E}_{pq}$ and $\hat{e}_{pqrs}$ are singlet one- and two-electron excitation operators.

\subsection{Quantum Linear Response}

After the UCC wave function optimization, the wave function can subsequently be used as the basis for evaluating molecular properties.
We will consider the quantum linear response (qLR) framework introduced in several recent works \cite{kumar2023quantum,ziems2024options,reinholdt2024subspace}, which for UCC wave functions is equivalent to qEOM\cite{taube2006new}.
Excitation energies are obtained by solving the generalized eigenvalue problem
\begin{equation}
    \mathbf{E}^{[2]} \mathbf{X}_k = E_{0k} \mathbf{S}^{[2]} \mathbf{X}_k~, 
\end{equation}
where the $\mathbf{E}^{[2]}$ is the Hessian and $\mathbf{S}^{[2]}$ is the metric, $E_{0k}$ is the transition energy, and $\mathbf{X}_k$ is the associated excitation vector. The matrix elements are defined as 

\begin{equation}
 \left(\begin{array}{cc}
\mathbf{A} & \mathbf{B}\\
\mathbf{B}^{*} & \mathbf{A}^{*}
\end{array}\right)\left(\begin{array}{c}
\mathbf{Y}_{k}\\
\mathbf{Z}_{k}
\end{array}\right)=E_{0k}\left(\begin{array}{cc}
\mathbf{\Sigma} & \mathbf{\Delta}\\
-\mathbf{\Delta}^{*} & \mathbf{-\Sigma}^{*}
\end{array}\right)\left(\begin{array}{c}
\mathbf{Y}_{k}\\
\mathbf{Z}_{k}
\end{array}\right).
\label{eq:qLR}
\end{equation}
\begin{equation}
    A_{ij}  = \left<\Psi_{\mathrm{UCC}}\left| \comm{\hat{G}_i ^{\dagger}}{\comm{\hat{H}}{\hat{G}_j}}\right|\Psi_{\mathrm{UCC}}\right>~,  \label{eq:Aij}
\end{equation}
\begin{equation}
    B_{ij}  = \left<\Psi_{\mathrm{UCC}}\left| \comm{\hat{G}_i}{\comm{\hat{H}}{\hat{G}_j}}\right|\Psi_{\mathrm{UCC}}\right>~,  
\end{equation}
\begin{equation}
    \Sigma_{ij}  = \left<\Psi_{\mathrm{UCC}}\left| \comm{\hat{G}_i ^{\dagger}}{\hat{G}_j}\right|\Psi_{\mathrm{UCC}}\right>~, 
\end{equation}
\begin{equation}
    \Delta_{ij}  = \left<\Psi_{\mathrm{UCC}}\left| \comm{\hat{G}_i}{\hat{G}_j}\right|\Psi_{\mathrm{UCC}}\right>~.  \label{eq:deltaij}
\end{equation}
with $\mathbf{Y}_k$ and $\mathbf{Z}_k$ being the excitation and de-excitation part of the qLR excitation vector, respectively.
We chose the same spin-adaption as in eqs. \eqref{eq:wf_par_1} and \eqref{eq:wf_par_2} for the $\hat{G}$ operators.
In the qLR method, these matrix elements are evaluated using a quantum device, followed by a classical diagonalization procedure to extract the excitation energies and vectors.

\subsection{Self-consistent operator manifold}
As suggested by \citeauthor{asthana2023quantum}\cite{asthana2023quantum}, one can introduce a so-called self-consistent operator manifold, in which the primitive excitation operators are transformed according to
\begin{align}
    \hat{G}^\mathrm{sc} = U(\boldsymbol{\theta}) \hat{G} U(\boldsymbol{\theta})^\dagger
\end{align}
Importantly, the self-consistent operators satisfy the killer condition, i.e., the operators cannot lead to deexcitation from the ground state and $\hat{G}^{^{\dagger}\mathrm{sc}}\ket{\Psi_{\mathrm{UCC}}}=0.$
With this choice of operator manifold, the matrix elements in Eqs. \eqref{eq:Aij}-\eqref{eq:deltaij} simplify considerably: the metric matrix becomes diagonal ($\boldsymbol{\Sigma}=\mathbf{1}$ and $\boldsymbol{\Delta}=\mathbf{0}$), and the $\mathbf{B}$ part of the electronic Hessian vanishes. 
This simplification is particularly useful since it removes the need for any (inevitably noisy) measurement of the metric.
Thus, the generalized eigenvalue problem reduces to a regular eigenvalue problem
\begin{equation}
    \mathbf{A}\mathbf{Y}_{k}=E_{0k}\mathbf{Y}_{k}.
    \label{eq:eigenvalue}
\end{equation}

\subsection{Hessian-vector products}
In the original formulations of qLR/qEOM, every matrix element has to be measured, leading to $O(N^8)$ matrix elements for a truncation of the operator expansion to singles and doubles. When aiming to treat larger molecular systems, such an explicit construction of the Hessian can become prohibitively expensive.
Fortunately, the solution to this problem is well-known from classical quantum chemistry, where, for example, Davidson-type methods\cite{DAVIDSON197587,olsen1988solution} are applied to tackle this issue.
Therein, the explicit evaluation of matrix elements is not performed; rather, only the result of multiplying the Hessian matrix with an arbitrary trial vector is needed, requiring only the measurement of $O(N^4)$ matrix elements per Hessian-vector product.
In the context of q-sc-LR, such methods have already been proposed by \citeauthor{kim2023two}\cite{kim2023two}, and \citeauthor{reinholdt2024subspace}\cite{reinholdt2024subspace}.
In Ref. \citenum{reinholdt2024subspace}, we formulated the Hessian-vector multiplication, $\boldsymbol{\sigma}_\mathbf{b} = \mathbf{A} \mathbf{b}$, with the trial vector $\mathbf{b}$ in terms of gradient evaluations in a finite-difference approach 
\begin{equation}
     \boldsymbol{\sigma}_\mathbf{b} = \mathbf{A} \mathbf{b} 
        \approx  \frac{\mathbf{g}\left(\boldsymbol{\theta}, + h\mathbf{b}\right) - \mathbf{g}\left(\boldsymbol{\theta}, - h\mathbf{b}\right) }{2h} , 
        \label{eq:hvp-grad}
    \end{equation}
where $h$ is a small positive parameter, and $\mathbf{g}$ is the gradient with respect to the parameters $\boldsymbol{\Theta} = \pm h \mathbf{b}$ in the state 
\begin{equation}
    \left|\Psi(\boldsymbol{\theta},\mathbf{\Theta})\right> = U(\boldsymbol{\theta})U(\mathbf{\Theta})\left|0\right>.
\end{equation}

In this work, we consider an alternative formulation based on superposition states.
Consider the contraction of the matrix elements $A_{ij}$ with a trial vector. Invoking commutator rules, the self-consistent operator definition, and the killer condition, we obtain
\begin{eqnarray}
\sigma_{b,i} \nonumber & = & \sum_{j}A_{ij}b_{j} \\ &=&\sum_{j}\left\langle  \nonumber \Psi_{\mathrm{UCC}}\left|\left[\hat{G}_{i}^{^{\dagger}\mathrm{sc}},\left[\hat{H},\hat{G}_{j}^{\mathrm{sc}}\right]\right]\right|\Psi_{\mathrm{UCC}}\right\rangle b_{j} \nonumber \\ &=& \nonumber \left\langle \Psi_{\mathrm{UCC}}\left|\left[\hat{G}_{i}^{^{\dagger}\mathrm{sc}},\left[\hat{H},\sum_{j}b_{j}\hat{G}_{j}^{\mathrm{sc}}\right]\right]\right|\Psi_{\mathrm{UCC}}\right\rangle \nonumber \\
 & =&  \left\langle \Psi_{\mathrm{UCC}}\left|G_{i}^{^{\dagger}\mathrm{sc}},\left[\hat{H},\hat{b}^{sc}\right]\right|\Psi_{\mathrm{UCC}}\right\rangle \nonumber \\ \nonumber &=& \left\langle \Psi_{\mathrm{UCC}}\left|G_{i}^{^{\dagger}\mathrm{sc}}\hat{H}\hat{b}^{\mathrm{sc}}-\hat{G}_{i}^{^{\dagger}\mathrm{sc}}\hat{b}^{\mathrm{sc}}\hat{H}\right|\Psi_{\mathrm{UCC}}\right\rangle \nonumber \\
 & = & \left\langle 0\left|\hat{G}_{i}^{^{\dagger}}U^{\dagger}\hat{H}U\hat{b}-\hat{G}_{i}^{^{\dagger}}\hat{b}U^{\dagger}\hat{H}U\right|0\right\rangle \nonumber \\
 & = & \left\langle 0\left|\hat{G}_{i}^{^{\dagger}}U^{\dagger}\hat{H}U\hat{b}\right|0\right\rangle -b_{i}E_0  \label{eq:hvp-super}
\end{eqnarray}
where we have introduced $\hat{b}=\sum_{j} b_j \hat{G}_j$ and $\hat{b}^{\mathrm{sc}}=\sum_{j} b_j \hat{G}_j^{\mathrm{sc}}$.
In the last line, we have assumed that the UCC residuals $\left\langle 0\left|\hat{G}_{i}^{^{\dagger}}\hat{G}_{j}U^{\dagger}\hat{H}U\right|0\right\rangle $ are zero for $i\neq j$. Although this is only strictly true when $\ket{\Psi}$ is an eigenfunction of $\hat{H}$, we assume this is a good approximation for the type of UCCSD wave functions we employ.

Inspired by earlier work by \citeauthor{nakanishi2019subspace}\cite{nakanishi2019subspace} and \citeauthor{asthana2023quantum}\cite{asthana2023quantum} for single elements of the Hessian, we can evaluate the transition matrix element between $\bra{0}\hat{G}_{i}^{^{\dagger}}$ and $\hat{b}\ket{0}$ on a quantum computer by measuring the expectation value of the ``superposition'' state $\hat{G}_i + \hat{b}$:
\begin{eqnarray}
A_{i+b,i+b} & = & \left\langle 0\left|\left(\left(\hat{G}_{i}+\hat{b}\right)^{\dagger}U^{\dagger}HU\left(\hat{G}_{i}+\hat{b}\right)\right)\right|0\right\rangle \\
 & = & \left\langle 0\left|\hat{G}_{i}^{\dagger}U^{\dagger}HU\hat{G}_{i}\right|0\right\rangle +\left\langle 0\left|\hat{b}^{\dagger}U^{\dagger}HU\hat{b}\right|0\right\rangle \nonumber\\
 & + & 2\left\langle 0\left|\hat{G}_{i}^{\dagger}U^{\dagger}HU\hat{b}\right|0\right\rangle\nonumber
\end{eqnarray}
This recasts our desired matrix element (the first term in eq. \eqref{eq:hvp-super}) as
\begin{equation}
    \left\langle 0\left|\hat{G}_{i}^{^{\dagger}}U^{\dagger}\hat{H}U\hat{b}\right|0\right\rangle = \frac{1}{2} \left( A_{i+b,i+b}   - A_{ii} - A_{b,b}\right). \label{eq:transition_matrix_element}
\end{equation}

Practically, this means preparing a set of ``simple'' superposition states of determinants based on excitation out of the reference state onto which the ground-state circuit $U$ is applied. 
For the state preparation, we use the approach by \citeauthor{mottonen2004transformation}\cite{mottonen2004transformation}, as implemented in PennyLane~\cite{pennylane}. 
The state vectors generated by these superpositions are sparse and only have a polynomial number of nonzero elements, which can be handled efficiently in terms of classical computing overhead. 
The states generated by $\hat{G}_i$ and  $\hat{b} = \sum_j\hat{G}_j$ are, in general, neither orthogonal nor normalized. This is problematic as state vectors need to be normalized to be representable on a quantum computer. We circumvent this problem by preparing a normalized state $\widetilde{\ket{\Phi}} = \frac{\ket{\Phi}}{\norm{\ket{\Phi}}}$ proportional to $\ket{\Phi}$ on the quantum computer. We then evaluate the expectation value of $\ket{\Phi}$ over an operator $\hat{O}$ via $\widetilde{\ket{\Phi}}$ re-scaled with the original norm:
\begin{equation}
    \bra{\Phi}\hat{O}\ket{\Phi} = \norm{\ket{\Phi}}^2 \widetilde{\bra{\Phi}}\hat{O}\widetilde{\ket{\Phi}}.
\end{equation}
The same essential procedure described above can also be applied to operators other than the Hamiltonian, e.g., the $\hat{E}_{pq}$, which we will later use to obtain transition densities.

\subsection{Polarizable Embedding}
In the polarizable embedding (PE) method, a molecular system is split into two parts: the QM region, which is treated using a quantum-mechanical description, and the environment, which is treated semi-classically through the combination of a distributed multipole expansion and a point induced dipole model.
The distributed multipole expansion accounts for the permanent electrostatic potential from the environment, while distributed polarizabilities in the environment model account for the mutual polarization between the environment and the QM region.
This is implemented by augmenting the regular gas-phase Hamiltonian with an additional embedding operator, $\hat{v}^\mathrm{emb}$:
\begin{equation}
    \hat{H}^{\mathrm{PE}}= \hat{H}^{\mathrm{vac}} + \hat{v}^\mathrm{emb} \label{eq:pe_hamiltonian}
\end{equation}
that is defined as
\begin{equation}
    \hat{v}^{\mathrm{emb}} = \hat{v}^\mathrm{es} + \hat{v}^\mathrm{ind}. \label{eq:emb_operator}
\end{equation}

The first term is the electrostatic operator, which contains the interaction between the electrons of the QM core and the static multipoles in the environment.
In the following, we will express the operators in terms of the atomic orbital (AO) basis matrix elements since that most closely follows our practical implementation. 
A matrix element of the electrostatic operator in the basis is 
\begin{equation}
    {v}_{\mu\nu}^{\mathrm{es}} = \sum_{s=1}^{N} \sum_{|k| = 0}^{K} \frac{(-1)^{|k|}}{k!} {{M}}_s^{(k)} {{t}}_{\mu\nu}^{(k)}(\mathbf{R}_s).
    \label{eq:v_es}
\end{equation}
Here, the indices $\mu,\nu$ refer to AO basis functions, and $s$ refers to a site in the environment.
We use a multi-index notation for the multipole order ($k$) and refer the reader to Ref. \citenum{olsen2011molecular} for further details on that topic. The distributed multipole moment tensors,
${{M}}_s^{(k)}$ are usually taken to be atomic charges, dipoles, and quadrupoles, and ${{t}}_{\mu\nu}^{(k)}(\mathbf{R}_s)$ is an integral over a $k$'th order interaction tensor $ {T}^{(k)}(\mathbf{R}) = \nabla^k \left(\frac{1}{R}\right)$ in the AO basis
\begin{equation}
    {{t}}_{\mu\nu}^{(k)}(\mathbf{R}_s) = -\int \chi_\mu(\mathbf{r}) {T}^{(k)}(\mathbf{R}_s) \chi_\nu(\mathbf{r}) \mathrm{d}\mathbf{r}.
\end{equation}
One can consider an environment description that includes only static multipoles, $\hat{v}^{\mathrm{emb}} = \hat{v}^\mathrm{es}$ (no induction operator term), which we will refer to as "static" in the following sections.

The second term in eq. \eqref{eq:emb_operator} is the induction operator, which contains the interaction between the induced dipoles of the environment and the electrons of the QM region. A matrix element of the induction operator in the AO basis is
\begin{equation}
    {v}_{\mu\nu}^{\mathrm{ind}} (\mathbf{D})= - \sum_{s=1}^{N} \sum_{\alpha=x,y,z} {\mu}^{\mathrm{ind}}_{s,\alpha}(\mathbf{D}) {t}_{\mu\nu,\alpha}^{(1)} (\mathbf{R}_{s}). \label{eq:induction_operator}
\end{equation}
The induction operator depends on the QM electronic density ($\mathbf{D}$), and on the induced dipoles of the environment, which are obtained from

\begin{equation}
    \boldsymbol{\mu}^{\mathrm{ind}}_s =  \boldsymbol{\alpha}_s \left( \mathbf{F}^{\mathrm{rhs}} + \mathbf{F}^{\mathrm{ind}} \right) =   \boldsymbol{\alpha}_s \left( \mathbf{F}^{\mathrm{mul}} + \mathbf{F}^{\mathrm{nuc}} + \mathbf{F}^{\mathrm{el}} + \sum_{s'\neq s} \mathbf{T}^{(2)}_{ss'} \boldsymbol{\mu}^{\mathrm{ind}}_{s'}  \right), \label{eq:mu_ind}
\end{equation}
where the ``right-hand-side'' field, $\mathbf{F}^{\mathrm{rhs}}$, is the sum of the fields from the static multipoles in the environment, $\mathbf{F}^{\mathrm{mul}}$, the fields from the nuclei of the QM region, $\mathbf{F}^{\mathrm{nuc}}$, and the fields from the electrons of the QM region, $\mathbf{F}^{\mathrm{el}}$.  
The fields from the QM electrons are obtained as
\begin{equation}
    F^{\mathrm{el}}_{\alpha}(\mathbf{R}_s) = \sum_{\mu\nu}  D_{\mu\nu} t^{(1)}_{\mu\nu,\alpha} (\mathbf{R}_s), \label{eq:qm_field}
\end{equation}
For the sake of brevity, we leave out explicit expressions for the fields from the nuclei and multipoles and refer to Ref. \citenum{stone2013theory} for explicit equations.  
As the induced moments appear on both sides of the equation via the induced field definition of eq. \eqref{eq:mu_ind}, we can recast it to a linear system of equations:
\begin{equation}
    \textbf{B} \boldsymbol{\mu}^{\mathrm{ind}} = \textbf{F}^{\mathrm{rhs}} \label{eq:induction_equations}.
\end{equation}
where
\begin{equation}
\mathbf{B}=\left(\begin{array}{cccc}
\boldsymbol{\alpha}_{1}^{-1} & -\mathbf{T}_{12}^{\left(2\right)} & \cdots & -\mathbf{T}_{1N}^{\left(2\right)}\\
-\mathbf{T}_{21}^{\left(2\right)} & \boldsymbol{\alpha}_{2}^{-1} & \cdots & -\mathbf{T}_{2N}^{\left(2\right)}\\
\vdots & \vdots & \ddots & \vdots\\
-\mathbf{T}_{N1}^{\left(2\right)} & -\mathbf{T}_{N2}^{\left(2\right)} & \cdots & \boldsymbol{\alpha}_{N}^{-1}
\end{array}\right).\end{equation}
The QM wave function depends on the induced dipoles obtained from solving eq. \eqref{eq:induction_equations}, which, in turn, depends on the QM electronic density.  
Thus, the ground-state PE description leads to a ``double-SCF'' procedure, in which during each wave function iteration, the polarization equations are solved with the most recent QM density, which is used to update the induction operator in the Hamiltonian. At convergence, the ground-state wave function and the induced dipoles are mutually and self-consistently polarized.

In practical terms, each optimization iteration involves first constructing the ground-state density $D_{pq} = \left<\Psi_{\mathrm{UCC}}(\boldsymbol{\theta})\left|\hat{E}_{pq}\right|\Psi_{\mathrm{UCC}}(\boldsymbol{\theta})\right>$, then transforming to the AO basis. The AO density matrix, together with the field integrals (also expressed in the AO basis), are used to evaluate the electric fields (eq. \eqref{eq:qm_field}) from the QM electrons. The electric fields enter on the right-hand side of the polarization equations.
After solving the polarization equations (eq. \eqref{eq:induction_equations}), which we do using an iterative solver, the induced dipole moments are obtained. These are used together with the field integrals to construct the induction operator (eq. \eqref{eq:induction_operator}). The induction operator is then transformed to the MO basis, followed by a fermion-to-qubit mapping (with Jordan-Wigner) to obtain the relevant qubit Hamiltonian.
One could, alternatively, have transformed the field integrals to the MO basis, avoiding the two basis transformations per iteration outlined above. However, the number of field integrals to be transformed ($3N$) can become quite considerable when tackling large MM environments, so we consider our approach preferable.

\subsection{Polarizable Embedding Hessian contributions}
Next, we will consider the effects introduced by the PE environment on the linear response equations. Following \citeauthor{olsen2010excited}\cite{olsen2010excited}, the embedding potential leads to contributions in the electronic hessian ($\mathbf{E}^{[2]}$), while the metric matrix ($\mathbf{S}^{[2]}$) is left unmodified.
We note in passing that there are also contributions to the property gradient through the so-called effective
external field (EEF) effects\cite{list2016local}, which practically speaking can be included by using a set of modified dipole integrals.
Including these contributions will correct intensities (e.g., oscillator strengths) but have no effect on the excitation energies, which is the main focus of this work.

There are two types of PE contributions to the electronic Hessian, which in terms of the $\sigma$-vectors can be formulated as
\begin{equation}
 \sum_{j }(A_{ij}^{\mathrm{PE}}b_j) = 
 \sigma_{b,i}^{\mathrm{gspol}} + \sigma_{b,i}^{\mathrm{dyn}} =-\left\langle 0\left|\left[\hat{G}_i,\left[\hat{b} ,\hat{v}_{\mathrm{PE}}^{0}\right] \right]\right|0\right\rangle
 -\left\langle 0\left|\left[\hat{G}_i,
 \hat{v}_{\mathrm{PE}}^{b}\right]\right|0\right\rangle,
\end{equation}
with $\hat{v}_{\mathrm{PE}}^{0} = \hat{v}^{\mathrm{es}} + \hat{v}^{\mathrm{ind}}[\mathbf{D}^{gs}]$.
The first term corresponds to the ground-state polarization and is included automatically by replacing the gas-phase Hamiltonian (see eq. \eqref{eq:Aij}) with the PE Hamiltonian (eq. \eqref{eq:pe_hamiltonian}) when evaluating the Hessian matrix elements (or in the construction of $\sigma$-vectors). Including only this term in the linear response treatment will be called "gspol" in the following sections.
The second term, sometimes called the dynamical response term, accounts for the environment's response to the perturbation, i.e., to the transition density from the QM region. The operator involved is defined as
\begin{equation}
    \hat{v}_{\mathrm{PE}}^{b}=-\sum_{s=1}^{S}\boldsymbol{\mu}_{s}^{\mathrm{ind}}\left(\tilde{\mathbf{F}}^{b}\right)\hat{\mathbf{T}}_{s}^{\left(1\right)}
\end{equation}
with the field being
\begin{equation}
    \tilde{\mathbf{F}}_{s}^{b}=\left\langle 0\left|\left[\hat{b},\hat{\mathbf{T}}_{s}^{\left(1\right)}\right]\right|0\right\rangle,
    \label{eq:transition_field}
\end{equation}
and the $\hat{\mathbf{T}}_{s}^{\left(1\right)}$ being defined as
\begin{equation}
    \hat{\mathbf{T}}_{s}^{\left(1\right)} = \sum_{pq} t^{(1)}_{pq}(\mathbf{R}_s) \hat{E}_{pq}.
\end{equation}
Including this term gives the full linear response from the polarizable environment and will be called ``dynpol'' in the following.

In practical terms, this means that the dynamical response contribution to the Hessian can be evaluated in a similar manner to a property gradient (over the $\hat{v}_{\mathrm{PE}}^{b}$ operator).
We construct the transition fields in \eqref{eq:transition_field} by building transition densities with the technique outlined in eq. \eqref{eq:transition_matrix_element}, i.e., we form superposition states according to the trial vectors, which allows us to evaluate the transition matrix element $D^{tr,b}_{pq}=\left<0\left|\hat{E}_{pq}\right|b\right>$.
Note that the dynamical response term leads to contributions in both the $\mathbf{A}$ and $\mathbf{B}$ blocks in equation \eqref{eq:qLR}, even when using the self-consistent operator manifold. The contributions have the structure 
\begin{equation}
    \mathbf{E}_\text{dyn}^{\left[2\right]}=\left(\begin{array}{cc}
\mathbf{A}_\text{dyn} & -\mathbf{A}_\text{dyn}\\
-\mathbf{A}_\text{dyn} & \mathbf{A}_\text{dyn}
\end{array}\right).
\end{equation}
This means that the simplification from a generalized eigenvalue problem to a regular eigenvalue problem, as discussed previously for a molecule in isolation,  is no longer possible. However, considering the usefulness of this simplification, we will investigate the validity of applying a Tamm-Dancoff-like approximation (i.e. ignoring the off-diagonal blocks), under which the simple structure of the regular eigenvalue problem is retained.

\subsection{Error mitigation}
When targeting real quantum hardware, an important consideration is that currently available devices are imperfect and subject to hardware noise. Such noise introduces systematic bias to any quantum computation, which can negatively impact the quality of the obtained results. 
In the near term, one option for dealing with this problem is applying quantum error mitigation techniques, which seek to reduce the impact of hardware errors.
We use the ansatz-based read-out and gate error mitigation strategy (M0) proposed in Ref. \citenum{ziems2024understanding}, which is
inspired by conventional read-out error mitigation (REM)\cite{maciejewski2020mitigation}. Mitigated quasi-probability vectors are obtained by transforming the measured (raw) probability vector as 
\begin{equation}
    p_\text{mitigated} = M^{-1} p_\text{raw}.
    \label{eq:M0_mitigate}
\end{equation}
Here, $M_{ij} = \mathrm{Pr}(q_j|q_i)$ is the 
{\em confusion} matrix, the elements of which are defined as the probability of measuring a bitstring $q_j$ when preparing a circuit to produce bitstring $q_i$.
In conventional REM, the confusion matrix encodes only read-out errors.
However, the M0 approach also corrects gate errors by measuring the elements of $M$ for a circuit that includes the ansatz $U(\boldsymbol{\theta})$ with $\boldsymbol{\theta}=\mathbf{0}$. 
This allows for correcting gate errors in the ansatz, which is particularly important for two-qubit gates such as CNOT gates.

We also investigate the add-and-subtract Clifford-based gate error mitigation (CBGEM) suggested by \citeauthor{motta2023quantum}\cite{motta2023quantum} This method starts out in a similar spirit to the ansatz-based read-out and gate error mitigation strategy by executing and classically simulating circuits with $\boldsymbol{\theta=\mathbf{0}}$ to encode gate error. Such circuits are classically tractable as they are completely Clifford-based. Next, they correct expectation values according to
\begin{equation}
    \left<P(\boldsymbol{\theta})\right>_\text{mitigated} = \left<P(\boldsymbol{\theta})\right>_\text{raw} + \left(\left<P(\mathbf{0})\right>_\text{ideal}  -  \left<P(\mathbf{0})\right>_\text{raw}\right). \label{eq:expectation_mitigate}
\end{equation}
Thus, this method corrects for gate error in individual expectation values rather than applying gate and read-out mitigation to the probability vectors for a given ansatz. 

\section{Computational Details}\label{sec:computational_details}
We have implemented the PE model in the codebase described in Ref. \citenum{reinholdt2024subspace}, available on \url{https://github.com/HQC2/subspace-response/}. 
Our implementation uses Pennylane\cite{pennylane} (version 0.38.0) for quantum circuit simulation with the Pennylane-Lightning state-vector simulator. 
For the simulations including simulated hardware noise, we used the Qiskit Aer quantum simulator\cite{qiskit2024}.
All PE-related integrals are obtained using PySCF\cite{pyscf}, which is also used to obtain the initial Hartree-Fock (HF) wave function and molecular orbitals. We include PE effects in the reference HF wave function through the CPPE library\cite{scheurer2019cppe}.
The fermionic one- and two-electron integrals are Jordan-Wigner-mapped using OpenFermion\cite{mcclean2020openfermion}. 
Ground-state UCCSD wave functions are optimized through minimization with the SLSQP\cite{slsqp} optimizer available in Scipy\cite{virtanen2020scipy}. 
Reference HF, CASCI, and CCSD calculations are carried out with the Dalton program\cite{daltonpaper}.
We note that the PE model is not implemented for CI wave functions in Dalton; it is only available through PE-MCSCF. The PE-CASCI results were obtained by converging a HF calculation and saving the orbitals to disk, then starting a PE-MCSCF calculation from those orbitals, and marking all orbitals as frozen, which removes all orbital rotation parameters from the wave function, yielding the desired PE-CASCI wavefunction.
The ``active-space'' CCSD calculations were obtained by freezing both occupied and virtual orbitals until the desired active space was recovered.

A snapshot of para-nitroaniline (pNA) in water was obtained from Ref. \citenum{reinholdt2021fast}, keeping the 80 nearest water molecules (based on the distance to any atom in pNA) in the embedding description. We used the PyFraME library\cite{pyframe} to generate the embedding potential, which contained atomically distributed charges, dipoles, quadrupoles, and anisotropic polarizabilities derived using the Loprop\cite{gagliardi2004local} method based on CAM-B3LYP\cite{yanai2004new}/loprop-aug-cc-pVTZ calculations in Dalton\cite{daltonpaper}, version 2020.1. 

Snapshots of butadiene in water were obtained from classical molecular dynamics simulations. The solvated geometry of butadiene in a box with 7000 water molecules was obtained from Ref. \citenum{van2023embedding}. A 100 ns production MD simulation was performed using the Gromacs program\cite{abraham2015gromacs}, version 2023. We used a $2\,$fs timestep, and long-range electrostatics were treated with the particle-mesh Ewald method\cite{darden1993particle}, with a short-range cutoff of $12\,$\AA. Bonds involving hydrogens were constrained using LINCS\cite{hess1997lincs}. The temperature was controlled by a Bussi-Donadio-Parrinello thermostat\cite{bussi2007canonical} towards $298.15\,$K, while the pressure was controlled towards 1 bar by a C-rescale\cite{crescale} barostat. Snapshots were extracted every $1\,$ns (for a total of 100 snapshots).
For each snapshot, the 80 nearest water molecules to the butadiene solute were retained, and embedding potentials were generated as described above.

\section{Results and Discussion}

\subsection{Implementation validation}
\begin{figure}
    \centering
    \includegraphics[width=0.5\linewidth]{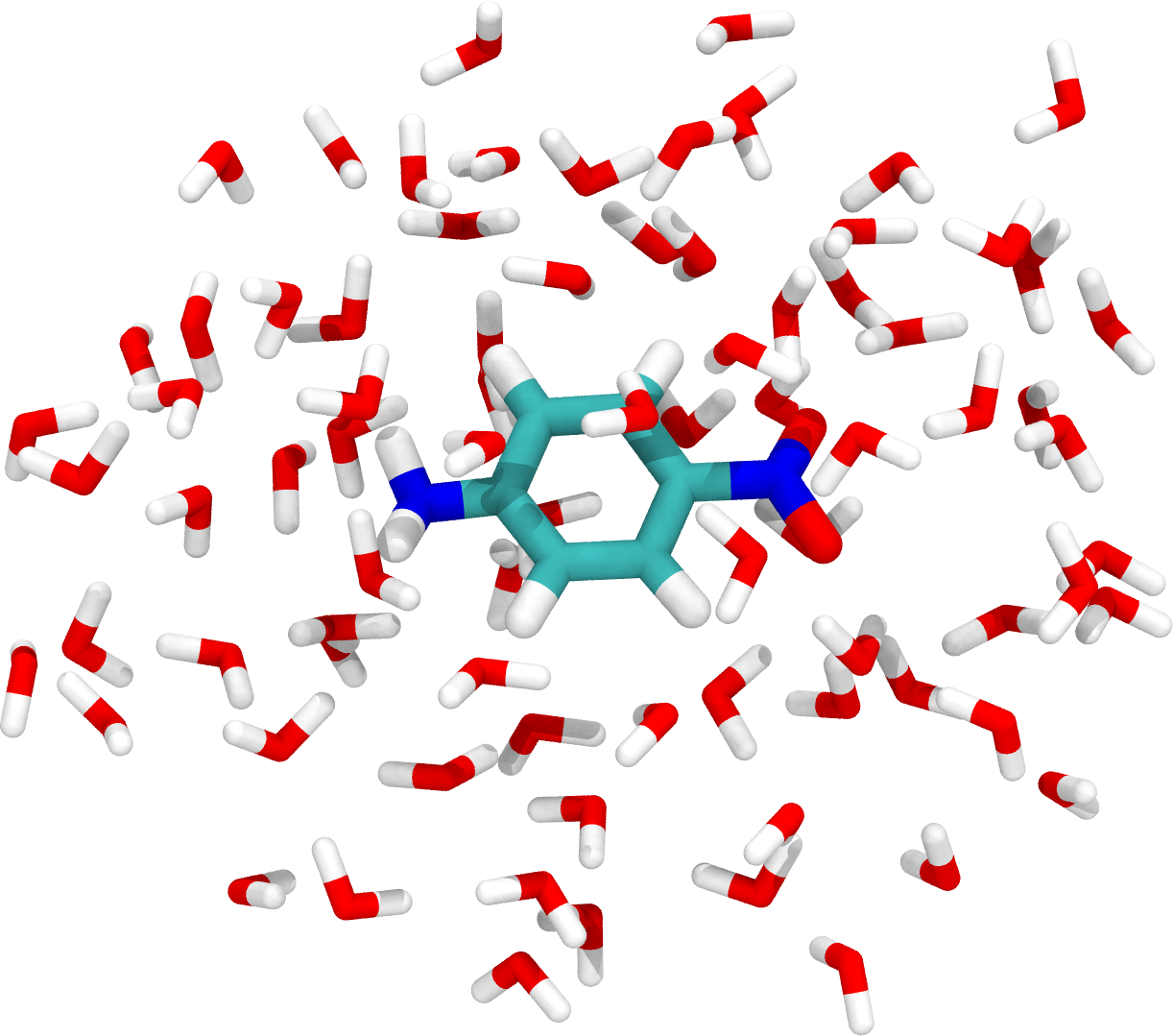}
    \caption{A snapshot of para-nitroaniline embedded in 80 water molecules. The structure was obtained from the MD simulation described in Ref. \citenum{reinholdt2021fast}.}
    \label{fig:pna_snapshot}
\end{figure}

We first validate our UCCSD implementation of PE-LR by comparing the excitation energies to reference classical PE-CASCI calculations. 
In Table \ref{tab:pelr-validation}, we report excitation of the para-nitroaniline (pNA) molecule solvated in water (see Figure \ref{fig:pna_snapshot}) with a 6-31+G* basis set and a (2,2) active space, using the HF HOMO and LUMO orbitals. With this active space, UCCSD will be equivalent to CASCI. There are two singlet excitations, one with predominantly single excitation character (an intramolecular charge-transfer excitation) with a large oscillator strength ($f=0.76$ with PE-LR) and a higher-lying double excitation with near-zero transition strength.
We find that our UCCSD implementation matches the reference CASCI results well (to within about $10^{-4}\,$eV).
The inclusion of the environment leads to significant changes in the excitation energies. For the lowest excitation, the electrostatic interaction lowers the excitation energy by about $0.64\,$eV, while ground-state polarization and dynamic response contributions further contribute with redshifts of $0.26$ and $0.12\,$eV, respectively. The second excitation energy (a double excitation) has a large blueshift of $0.54\,$eV upon including the electrostatic interaction while accounting for the ground state polarization red-shifts the excitation energy by about $0.70\,$eV. The dynamical response contribution is almost non-existent for this transition since the transition density is near-zero for the double excitation.
Note that these (2,2) active space simulations are a test model for our quantum approach. The large solvation shifts are likely not entirely physical but rather related to using the small (2,2) active space. For example, \citeauthor{nanda2018effect}\cite{nanda2018effect} observed much smaller solvation shifts of pNA in water of around $0.1\,$eV with (full-QM) EOM-EE-CCSD calculations, albeit on rather small water clusters (pNA in 6 waters). Similarly, we find a solvation shift of $0.04\,$eV for the first excitation at the HF level for the present snapshot of pNA in water. The apparently large solvation effects observed in the (2,2) active spaces are related to significant changes in the frontier orbitals upon solvation and the fact that the excitations are \emph{only} expanded in those orbitals. 
Nevertheless, this system serves as a decent testing ground for validating our PE implementation.

\begin{table}
    \centering
\begin{tabular}{|r|rrrrr|}
\hline
Excitation energies (eV) & CASCI & CASCI & CASCI & CASCI & \tabularnewline
State  & Gas-phase & Static & gspol & LR & \tabularnewline
1 & 5.66726 & 5.02318 & 4.76075 & 4.64190 & \tabularnewline
2 & 16.09191 & 16.62799 & 15.93253 & 15.93219 & \tabularnewline
\hline
 Excitation energies (eV) & UCCSD & UCCSD & UCCSD & UCCSD & UCCSD\tabularnewline
State  & Gas-phase & Static & gspol & LR & TDA\tabularnewline
1 & 5.66726 & 5.02321 & 4.76077 & 4.64191 & 4.64340\tabularnewline
2 & 16.09195 & 16.62813 & 15.93265 & 15.93232 & 15.93232\tabularnewline
\hline
Errors (eV) & UCCSD & UCCSD & UCCSD & UCCSD & UCCSD\tabularnewline
State  & Gas-phase & Static & gspol & LR & TDA$^{a}$ \tabularnewline
1 & 2.50$\times10^{-6}$ & 2.50$\times10^{-5}$ & 1.55$\times10^{-5}$ & 1.25$\times10^{-5}$ & 1.49$\times10^{-3}$\tabularnewline
2 & 4.40$\times10^{-5}$ & 1.45$\times10^{-4}$ & 1.27$\times10^{-4}$ & 1.27$\times10^{-4}$ & 2.50$\times10^{-6}$\tabularnewline
\hline
\end{tabular}
    \caption{Excitation energies of pNA/6-31+G* in water with various environment descriptions. Both CASCI and UCCSD wavefunctions are expanded in a (2,2) active space. Errors relative to the CASCI reference excitation energies are given in the lower columns. $^a$ Error relative to the UCCSD full linear response (PE-LR) result.}
    \label{tab:pelr-validation} %
\end{table}

\subsection{Effect of the Tamm-Dancoff-like approximation}

\begin{figure}
    \centering
    \includegraphics[width=0.5\linewidth]{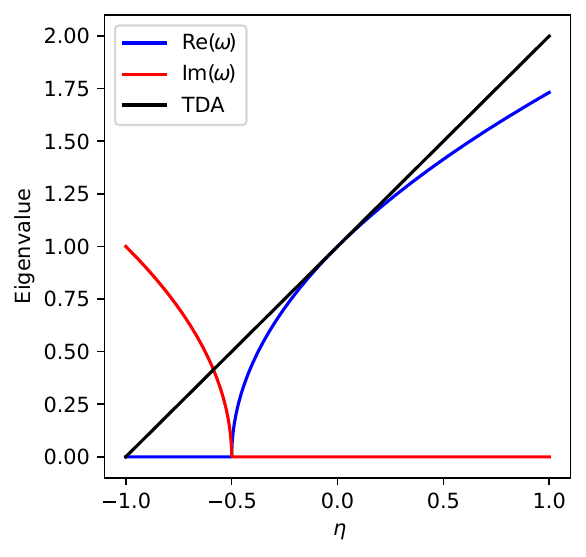}
    \caption{The eigenvalue ($\omega$) from the TDA approximation compared to the full model generalized eigenvalue problem as a function of the interaction strength $\eta$. Only the positive eigenvalue is shown.}
    \label{fig:tda-model}
\end{figure}

Next, we consider the effects of invoking the before-mentioned Tamm-Dancoff-like approximation (TDA) in the environmental coupling. As seen in Table \ref{tab:pelr-validation}, the effects of applying TDA are relatively minor (around $10^{-3}\,$eV), even with the relatively large transition density associated with the lowest excited state for pNA. At first glance, this finding might seem somewhat surprising since the TDA neglects a contribution to the Hessian that is of equal magnitude to the dynamical response contribution itself. 
To understand this observation, we consider the eigenvalues from the simplified model system  
\begin{equation}
    \left(\mathbf{E}_\text{stat}^{\left[2\right]}+ \mathbf{E}_\text{dyn}^{\left[2\right]}\right)\mathbf{X}=\omega \mathbf{S}^{\left[2\right]}\mathbf{X}
\end{equation}
with
\begin{equation}
\mathbf{E}_\text{stat}^{\left[2\right]}=\left(\begin{array}{cc}
1 & 0\\
0 & 1
\end{array}\right),\quad \mathbf{E}_\text{dyn}^{\left[2\right]}=\eta\left(\begin{array}{cc}
1 & -1\\
-1 & 1
\end{array}\right),\quad \mathbf{S}^{\left[2\right]}=\left(\begin{array}{cc}
1 & 0\\
0 & -1
\end{array}\right).
\end{equation}
The parameter $\eta$ controls the relative magnitude between the static and dynamic contributions to the total Hessian in the model problem. Using the TDA approximation, the eigenvalues become $\pm(1+\eta)$. 
As shown in Figure \ref{fig:tda-model}, the TDA approximation is reasonable when $\eta$ is small (the region where the TDA line is tangent to the blue one), i.e., when the magnitude of $\mathbf{E}^{\left[2\right]}_\text{stat}$ is large compared to $\mathbf{E}^{\left[2\right]}_\text{dyn}$. This is the case for real-world systems, even for relatively large transition densities.
The largest contributions to $\mathbf{E}^{\left[2\right]}$ are by far from the gas-phase Hamiltonian, followed by the electrostatic and ground-state induction contributions, with the dynamic response contributions being somewhat smaller in magnitude.
For example, for the pNA molecule used in table \ref{tab:pelr-validation}, $||\mathbf{E}^{[2]}_\text{stat}|| = 0.611$ a.u., while $||\mathbf{E}^{[2]}_\text{dyn}|| = 4.32\times 10^{-3}$ a.u.
Considering the relatively small errors associated with applying TDA to the dynamical polarization contribution and the resulting decreased computational complexity of the calculations, we will apply this approximation in all the following examples.

\subsection{Absorption energy of butadiene in water}

Now, we turn to a more ``real-world'' example, modeling the absorption energy of the butadiene molecule solvated in water. We will use a (4,4) active space, i.e., the $\pi$ space, along with the 6-31+G* basis set.
Table \ref{tab:butadiene_excitation} reports the average excitation energies computed with CASCI, CCSD, and UCCSD across a 100-snapshot trajectory, using the CASCI (4,4) results as a reference. As a point of comparison, we also include full-space CCSD results, which can indicate the impact of missing dynamic correlation from the use of a small active space.
We consider the error in the average excitation energies (subject to error cancellation since errors can have both positive and negative signs).
We find that both CCSD and UCCSD match the reference CASCI results quite well, with errors in the milli-eV range. For this system, UCCSD slightly outperforms conventional CCSD, particularly for the lowest excited state, with absolute errors in the excitation energy smaller by a factor of about 3. The larger errors are due to a systematic red-shift of the lowest excitation energy with CCSD. However, in absolute terms, the errors with CCSD are nevertheless quite small, especially considering the effects of missing correlation treatment with the small active space.
In particular, comparing the CCSD (4,4) active-space results to the full-space CCSD transition energies, we can estimate that the impact of missing dynamical correlation amounts to a red-shift of about $0.4\,$eV, which is certainly much more significant than any observed differences in excitation energies between the regular and unitary coupled-cluster approaches.

\begin{table}
    \centering
\begin{tabular}{|c|ccccc|}
\hline 
$\Delta E$ (eV) & HF & CASCI (4,4) & UCCSD (4,4) & CCSD (4,4) & CCSD\tabularnewline
\hline 
1 & 5.4817 & 6.3728 & 6.3746 & 6.3672 & 5.9890\tabularnewline
2 & 6.5865 & 7.0691 & 7.0704 & 7.0680 & 6.6449\tabularnewline
\hline 
Errors (eV) &  &  & UCCSD (4,4) & CCSD (4,4) & Correlation\tabularnewline
\hline 
 1&  &  & 0.0018 & $-$0.0056 & $-$0.3782\tabularnewline
 2&  &  & 0.0013 & $-$0.0011 & $-$0.4231\tabularnewline
\hline 
\end{tabular}
    \caption{Average excitation energies (top) and excitation energy errors (bottom) of butadiene/6-31+G* in water across a 100-snapshot trajectory. Errors for UCCSD (4,4) and CCSD (4,4) are computed relative to the CASCI (4,4) reference. The correlation correction to the excitation energy is estimated as the difference between the CCSD (4,4) and full-space CCSD results.}
    \label{tab:butadiene_excitation}
\end{table}


\subsection{Effects of shot noise}

\begin{figure}
    \centering
    \includegraphics[width=0.8\linewidth]{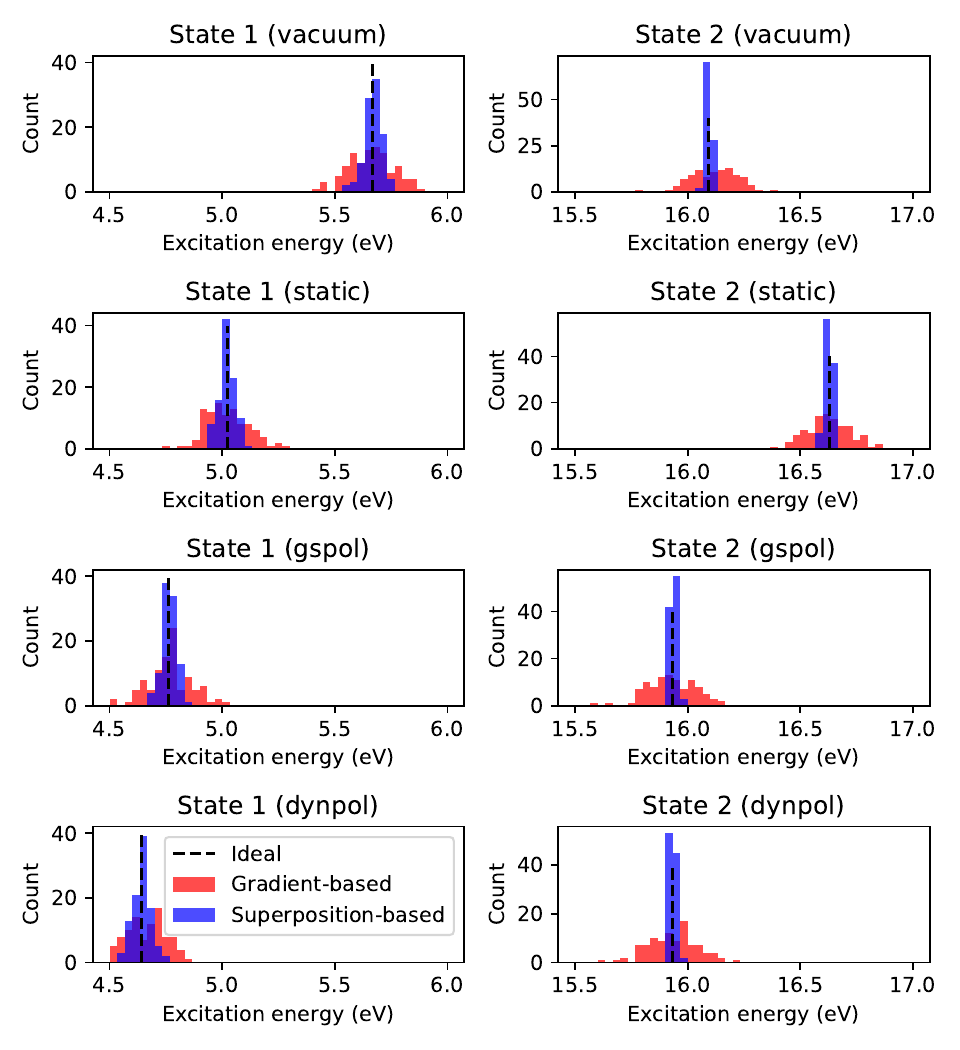}
    \caption{Excitation energy distributions from sample excitation energy calculations ($N=100$) due to shot noise errors in the hessian-vector evaluation for the first- and second excited state of pNA(2,2)/6-31+G* in water. The environment description varies from none (gas phase), electrostatic (charges, dipoles, and quadrupoles), polarization in the ground state (termed gspol), and full dynamical polarization (termed dynpol).   We compare Hessian-vector products evaluated with a gradient-based formulation (7-point, $h=0.5$) to a superposition-state-based formulation.}
    \label{fig:hvp-comparison}
\end{figure}

As previously mentioned, we have considered two approaches to computing Hessian-vector products, namely a gradient-based formulation (Eq. \eqref{eq:hvp-grad}) and a formulation based on superposition states (Eq. \eqref{eq:hvp-super}). In Figure \ref{fig:hvp-comparison}, we compare the shot-noise sensitivity of these two approaches using a fixed shot budget ($10^5$ shots). We use the same pNA molecule previously discussed (see Fig. \ref{fig:pna_snapshot}).
Our main conclusion from this analysis is that the new approach based on superposition states (blue bars) leads to significantly lower statistical noise. 
For example, the mean absolute deviations (MAD) of the fully polarizable environment (including dynamical response) to the reference (noise-free) is $0.031$ and $0.012\,$eV for the first and second excited states of the superposition state approach, while it is $0.082$ and $0.088\,$eV for the gradient-based approach. This amounts to a reduction in the MAD by a factor 2.5--7.3.

Regarding the effect of changing the environment description (different rows), we find that introducing the polarizable environment does not significantly impact the statistical errors in the excitation energies.
For example, for the superposition states approach (blue bars), the standard deviations are comparable for vacuum calculations (0.038 eV), static (0.036 eV), ground-state polarization (0.033 eV), and full dynamical polarization (0.041 eV).
This observation is consistent with the above analysis concerning the magnitude of the Hessian contributions, namely that for many typical molecular systems, 
the largest contributions to the Hessian are from the gas-phase Hamiltonian, followed by the electrostatic and ground-state induction contributions, with the dynamic response contributions being the smallest in magnitude.
It is evidently possible to capture the effects of changing the environment description even in the presence of statistical noise.

\subsection{Effects of simulated hardware noise and mitigation strategies}

\begin{figure}
    \centering
    \includegraphics[width=0.8\linewidth]{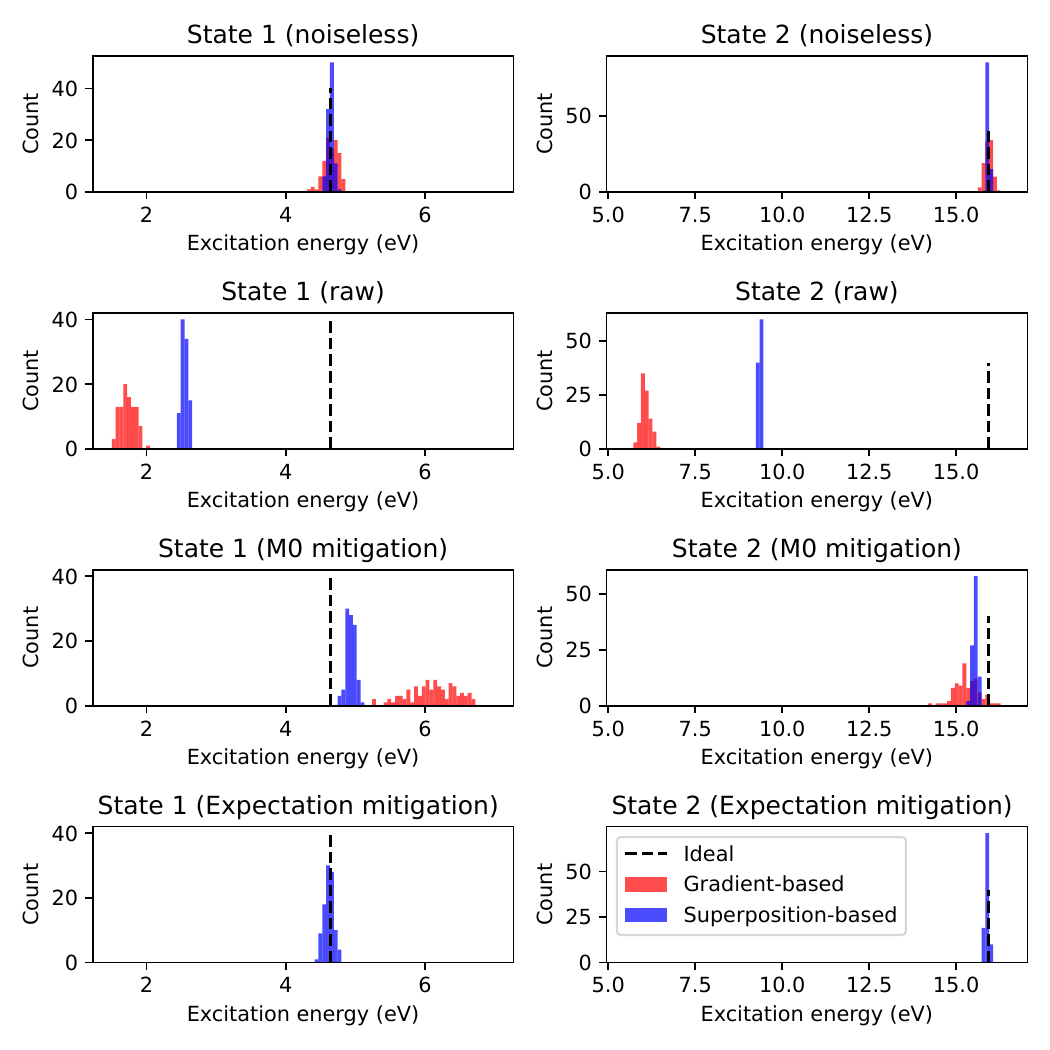}
    \caption{Excitation energy distributions from sample excitation energy calculations ($N=100$) due to shot noise and gate errors in the Hessian-vector evaluation for the first and second excited state of pNA(2,2)/6-31+G* in water. The environment description includes full dynamical polarization (dynpol).}
    \label{fig:errors-faketorino}
\end{figure}

In addition to simple statistical (shot) noise, current quantum devices are imperfect and also exhibit hardware noise, which may introduce significant bias to computed expectation values and any properties derived thereof.
We test the effects of such hardware noise in Figure \ref{fig:errors-faketorino}, using simulated hardware noise with the \texttt{faketorino} noise model\cite{qiskit2024}, which is designed to mimic the hardware noise present on the \texttt{ibm\_torino} quantum device.
Comparing the noiseless results to the raw (unmitigated) excitation energies, we find that hardware noise significantly impacts the quality of the computed results.
For the lowest excited state with the superposition-state-based hessian-vector product, the hardware noise introduces a massive $2.1\,$eV redshift in the excitation energy and an even larger error of $6.6\,$eV for the second excited state.

In the following, we will consider two approaches for handling the significant bias introduced by the hardware noise: ansatz-based read-out and gate error mitigation (M0) and add-and-subtract Clifford-based gate error mitigation (CBGEM). 
In M0 mitigation, additional execution precedes the calculation by measuring the confusion matrix $M$ of the null Ansatz circuit. This approach is compatible with both hessian-vector product formulations considered in this work.
After the confusion matrix is available, each circuit evaluation is as expensive as a regular, unmitigated execution in terms of quantum resources, albeit with some classical post-processing overhead.
In CBGEM, any expectation value is evaluated \emph{twice} on the quantum device, with either the actual circuit parameters or with all parameters set equal to zero, after which the corrected expectation value is obtained from Eq. \eqref{eq:expectation_mitigate}. Note that this approach is only compatible with the Hessian-vector formulation based on superposition states.

From Figure \ref{fig:errors-faketorino}, we find that both mitigation methods are reasonably effective in correcting hardware errors. For the M0 mitigation, we obtain errors (MAD) of $0.29$ and $0.39\,$eV for the first/second excited state with the superposition-based approach, while the gradient-based approach shows larger errors of 1.49 and 0.63 eV. 
We note that, as implemented, the M0 mitigation is slightly disadvantaged since we only measure the confusion matrix on the Ansatz, i.e.\ the UCCSD-based circuit, not the state-preparation circuit, as the latter will vary depending on which trial vector state is prepared. For the present example, this means that only 64 of the total 96 CNOT gates (about 70 \%) of CNOT gates are mitigated. It will be part of future work to extend the M0 to include the state-preparation circuit part. 

Even better results are obtained for the expectation-value-based mitigation with the superposition-based approach, where we obtain errors of $0.065$ and $0.076\,$eV for the first and second excited states, respectively. In both cases, the excitation energies are systematically under-estimated (by $0.03$ and $0.07\,$eV).
The CBGEM mitigation approaches can be expected to work well when the target state is close to being a Clifford circuit, i.e.\ the HF state, as all error mitigation is based on knowing the exact error for the Clifford version of a circuit expectation value. The present system (a pNA molecule in a small (2,2) active space) is not particularly multiconfigurational, and the overlap with the HF state is significant ($\left<\mathrm{HF}|\Psi_\mathrm{UCC}\right> = 0.99877$). This means that although the performance of the mitigated results appears promising, it is still to be investigated further whether such favorable results remain when targeting more challenging and multiconfigurational molecular systems.

\section{Conclusion}\label{sec:conclusion}
We have implemented the polarizable embedding model in combination with q-sc-LR, which is exemplified for UCCSD wave functions. This method allows computing excitation energies of molecular systems in solution using the Davidson method for the response equations. Moreover, we introduce a new superposition-state-based approach for computing the Hessian-vector products, which are at the core of such matrix-free methods.
The polarizable environment leads to several new contributions to the electronic Hessian. 
The so-called dynamic polarization contribution, in particular, introduces changes to the electronic Hessian that formally prevent the q-sc-LR equations from being solved as a simple eigenvalue problem but rather require a full solution of a generalized eigenvalue problem. 
We demonstrate that the original eigenvalue structure can safely be recovered by applying a Tamm-Dancoff-like approximation in the environment coupling without significant errors.
Using butadiene in water as an example, we show that the PE-UCCSD and conventional PE-CCSD perform similarly and that missing correlation corrections from using limited active spaces are significantly larger than any differences between the two methods.
Using the para-nitroaniline molecule in water, we investigate the effects of noise when computing excitation energies in solution. In terms of pure statistical noise, our proposed method performs acceptably well, and we demonstrate that the introduction of the polarizable environment does not significantly impact the sensitivity to statistical noise.
Concerning (simulated) hardware noise, we find that the quality of computed excitation energies is significantly impacted, but simple mitigation approaches reduce systematic errors enough to recover useful results.

\begin{acknowledgement}
We acknowledge the financial support of the Novo Nordisk Foundation for the focused research project \textit{Hybrid Quantum Chemistry on Hybrid Quantum Computers} (HQC)$^2$, grant number NNFSA220080996.
\end{acknowledgement}


\bibliography{main}

\providecommand{\latin}[1]{#1}
\makeatletter
\providecommand{\doi}
  {\begingroup\let\do\@makeother\dospecials
  \catcode`\{=1 \catcode`\}=2 \doi@aux}
\providecommand{\doi@aux}[1]{\endgroup\texttt{#1}}
\makeatother
\providecommand*\mcitethebibliography{\thebibliography}
\csname @ifundefined\endcsname{endmcitethebibliography}
  {\let\endmcitethebibliography\endthebibliography}{}
\begin{mcitethebibliography}{77}
\providecommand*\natexlab[1]{#1}
\providecommand*\mciteSetBstSublistMode[1]{}
\providecommand*\mciteSetBstMaxWidthForm[2]{}
\providecommand*\mciteBstWouldAddEndPuncttrue
  {\def\EndOfBibitem{\unskip.}}
\providecommand*\mciteBstWouldAddEndPunctfalse
  {\let\EndOfBibitem\relax}
\providecommand*\mciteSetBstMidEndSepPunct[3]{}
\providecommand*\mciteSetBstSublistLabelBeginEnd[3]{}
\providecommand*\EndOfBibitem{}
\mciteSetBstSublistMode{f}
\mciteSetBstMaxWidthForm{subitem}{(\alph{mcitesubitemcount})}
\mciteSetBstSublistLabelBeginEnd
  {\mcitemaxwidthsubitemform\space}
  {\relax}
  {\relax}

\bibitem[Kitaev(1995)]{kitaev1995quantum}
Kitaev,~A.~Y. Quantum measurements and the Abelian stabilizer problem.
  \emph{arXiv preprint quant-ph/9511026} \textbf{1995}, \relax
\mciteBstWouldAddEndPunctfalse
\mciteSetBstMidEndSepPunct{\mcitedefaultmidpunct}
{}{\mcitedefaultseppunct}\relax
\EndOfBibitem
\bibitem[Aspuru-Guzik \latin{et~al.}(2005)Aspuru-Guzik, Dutoi, Love, and
  Head-Gordon]{aspuru2005simulated}
Aspuru-Guzik,~A.; Dutoi,~A.~D.; Love,~P.~J.; Head-Gordon,~M. Simulated quantum
  computation of molecular energies. \emph{Sci} \textbf{2005}, \emph{309},
  1704--1707\relax
\mciteBstWouldAddEndPuncttrue
\mciteSetBstMidEndSepPunct{\mcitedefaultmidpunct}
{\mcitedefaultendpunct}{\mcitedefaultseppunct}\relax
\EndOfBibitem
\bibitem[Nash \latin{et~al.}(2020)Nash, Gheorghiu, and Mosca]{nash2020quantum}
Nash,~B.; Gheorghiu,~V.; Mosca,~M. Quantum circuit optimizations for NISQ
  architectures. \emph{Quantum Sci. Technol.} \textbf{2020}, \emph{5},
  025010\relax
\mciteBstWouldAddEndPuncttrue
\mciteSetBstMidEndSepPunct{\mcitedefaultmidpunct}
{\mcitedefaultendpunct}{\mcitedefaultseppunct}\relax
\EndOfBibitem
\bibitem[Lau \latin{et~al.}(2022)Lau, Lim, Shrotriya, and Kwek]{lau2022nisq}
Lau,~J. W.~Z.; Lim,~K.~H.; Shrotriya,~H.; Kwek,~L.~C. NISQ computing: where are
  we and where do we go? \emph{AAPPS Bulletin} \textbf{2022}, \emph{32},
  27\relax
\mciteBstWouldAddEndPuncttrue
\mciteSetBstMidEndSepPunct{\mcitedefaultmidpunct}
{\mcitedefaultendpunct}{\mcitedefaultseppunct}\relax
\EndOfBibitem
\bibitem[Peruzzo \latin{et~al.}(2014)Peruzzo, McClean, Shadbolt, Yung, Zhou,
  Love, Aspuru-Guzik, and O’brien]{peruzzo2014variational}
Peruzzo,~A.; McClean,~J.; Shadbolt,~P.; Yung,~M.-H.; Zhou,~X.-Q.; Love,~P.~J.;
  Aspuru-Guzik,~A.; O’brien,~J.~L. A variational eigenvalue solver on a
  photonic quantum processor. \emph{Nat. Commun.} \textbf{2014}, \emph{5},
  4213\relax
\mciteBstWouldAddEndPuncttrue
\mciteSetBstMidEndSepPunct{\mcitedefaultmidpunct}
{\mcitedefaultendpunct}{\mcitedefaultseppunct}\relax
\EndOfBibitem
\bibitem[McClean \latin{et~al.}(2016)McClean, Romero, Babbush, and
  Aspuru-Guzik]{mcclean2016theory}
McClean,~J.~R.; Romero,~J.; Babbush,~R.; Aspuru-Guzik,~A. The theory of
  variational hybrid quantum-classical algorithms. \emph{New J. Phys.}
  \textbf{2016}, \emph{18}, 023023\relax
\mciteBstWouldAddEndPuncttrue
\mciteSetBstMidEndSepPunct{\mcitedefaultmidpunct}
{\mcitedefaultendpunct}{\mcitedefaultseppunct}\relax
\EndOfBibitem
\bibitem[Ollitrault \latin{et~al.}(2020)Ollitrault, Kandala, Chen, Barkoutsos,
  Mezzacapo, Pistoia, Sheldon, Woerner, Gambetta, and
  Tavernelli]{ollitrault2020quantum}
Ollitrault,~P.~J.; Kandala,~A.; Chen,~C.-F.; Barkoutsos,~P.~K.; Mezzacapo,~A.;
  Pistoia,~M.; Sheldon,~S.; Woerner,~S.; Gambetta,~J.~M.; Tavernelli,~I.
  Quantum equation of motion for computing molecular excitation energies on a
  noisy quantum processor. \emph{Phys. Rev. Research} \textbf{2020}, \emph{2},
  043140\relax
\mciteBstWouldAddEndPuncttrue
\mciteSetBstMidEndSepPunct{\mcitedefaultmidpunct}
{\mcitedefaultendpunct}{\mcitedefaultseppunct}\relax
\EndOfBibitem
\bibitem[Asthana \latin{et~al.}(2023)Asthana, Kumar, Abraham, Grimsley, Zhang,
  Cincio, Tretiak, Dub, Economou, Barnes, and Mayhall]{asthana2023quantum}
Asthana,~A.; Kumar,~A.; Abraham,~V.; Grimsley,~H.; Zhang,~Y.; Cincio,~L.;
  Tretiak,~S.; Dub,~P.~A.; Economou,~S.~E.; Barnes,~E. \latin{et~al.}  Quantum
  self-consistent equation-of-motion method for computing molecular excitation
  energies, ionization potentials, and electron affinities on a quantum
  computer. \emph{Chem. Sci.} \textbf{2023}, \emph{14}, 2405--2418\relax
\mciteBstWouldAddEndPuncttrue
\mciteSetBstMidEndSepPunct{\mcitedefaultmidpunct}
{\mcitedefaultendpunct}{\mcitedefaultseppunct}\relax
\EndOfBibitem
\bibitem[Kumar \latin{et~al.}(2023)Kumar, Asthana, Abraham, Crawford, Mayhall,
  Zhang, Cincio, Tretiak, and Dub]{kumar2023quantum}
Kumar,~A.; Asthana,~A.; Abraham,~V.; Crawford,~T.~D.; Mayhall,~N.~J.;
  Zhang,~Y.; Cincio,~L.; Tretiak,~S.; Dub,~P.~A. Quantum Simulation of
  Molecular Response Properties in the NISQ Era. \emph{J. Chem. Theory Comput.}
  \textbf{2023}, \emph{19}, 9136–9150\relax
\mciteBstWouldAddEndPuncttrue
\mciteSetBstMidEndSepPunct{\mcitedefaultmidpunct}
{\mcitedefaultendpunct}{\mcitedefaultseppunct}\relax
\EndOfBibitem
\bibitem[Ziems \latin{et~al.}(2024)Ziems, Kjellgren, Reinholdt, Jensen, Sauer,
  Kongsted, and Coriani]{ziems2024options}
Ziems,~K.~M.; Kjellgren,~E.~R.; Reinholdt,~P.; Jensen,~P.~W.; Sauer,~S.~P.;
  Kongsted,~J.; Coriani,~S. Which options exist for NISQ-friendly linear
  response formulations? \emph{J. Chem. Theory Comput.} \textbf{2024},
  \emph{20}, 3551--3565\relax
\mciteBstWouldAddEndPuncttrue
\mciteSetBstMidEndSepPunct{\mcitedefaultmidpunct}
{\mcitedefaultendpunct}{\mcitedefaultseppunct}\relax
\EndOfBibitem
\bibitem[Reinholdt \latin{et~al.}(2024)Reinholdt, Kjellgren, Fuglsbjerg, Ziems,
  Coriani, Sauer, and Kongsted]{reinholdt2024subspace}
Reinholdt,~P.; Kjellgren,~E.~R.; Fuglsbjerg,~J.~H.; Ziems,~K.~M.; Coriani,~S.;
  Sauer,~S.~P.; Kongsted,~J. Subspace methods for the simulation of molecular
  response properties on a quantum computer. \emph{J. Chem. Theory Comput.}
  \textbf{2024}, \emph{20}, 3729--3740\relax
\mciteBstWouldAddEndPuncttrue
\mciteSetBstMidEndSepPunct{\mcitedefaultmidpunct}
{\mcitedefaultendpunct}{\mcitedefaultseppunct}\relax
\EndOfBibitem
\bibitem[Jensen \latin{et~al.}(2024)Jensen, Kjellgren, Reinholdt, Ziems,
  Coriani, Kongsted, and Sauer]{jensen2024quantum}
Jensen,~P.~W.; Kjellgren,~E.~R.; Reinholdt,~P.; Ziems,~K.~M.; Coriani,~S.;
  Kongsted,~J.; Sauer,~S.~P. Quantum Equation of Motion with Orbital
  Optimization for Computing Molecular Properties in Near-Term Quantum
  Computing. \emph{J. Chem. Theory Comput.} \textbf{2024}, \emph{20},
  3613--3625\relax
\mciteBstWouldAddEndPuncttrue
\mciteSetBstMidEndSepPunct{\mcitedefaultmidpunct}
{\mcitedefaultendpunct}{\mcitedefaultseppunct}\relax
\EndOfBibitem
\bibitem[McClean \latin{et~al.}(2017)McClean, Kimchi-Schwartz, Carter, and
  De~Jong]{mcclean2017hybrid}
McClean,~J.~R.; Kimchi-Schwartz,~M.~E.; Carter,~J.; De~Jong,~W.~A. Hybrid
  quantum-classical hierarchy for mitigation of decoherence and determination
  of excited states. \emph{Phys. Rev. A} \textbf{2017}, \emph{95}, 042308\relax
\mciteBstWouldAddEndPuncttrue
\mciteSetBstMidEndSepPunct{\mcitedefaultmidpunct}
{\mcitedefaultendpunct}{\mcitedefaultseppunct}\relax
\EndOfBibitem
\bibitem[Colless \latin{et~al.}(2018)Colless, Ramasesh, Dahlen, Blok,
  Kimchi-Schwartz, McClean, Carter, de~Jong, and
  Siddiqi]{colless2018computation}
Colless,~J.~I.; Ramasesh,~V.~V.; Dahlen,~D.; Blok,~M.~S.;
  Kimchi-Schwartz,~M.~E.; McClean,~J.~R.; Carter,~J.; de~Jong,~W.~A.;
  Siddiqi,~I. Computation of molecular spectra on a quantum processor with an
  error-resilient algorithm. \emph{Phys. Rev. X} \textbf{2018}, \emph{8},
  011021\relax
\mciteBstWouldAddEndPuncttrue
\mciteSetBstMidEndSepPunct{\mcitedefaultmidpunct}
{\mcitedefaultendpunct}{\mcitedefaultseppunct}\relax
\EndOfBibitem
\bibitem[McClean \latin{et~al.}(2020)McClean, Jiang, Rubin, Babbush, and
  Neven]{mcclean2020decoding}
McClean,~J.~R.; Jiang,~Z.; Rubin,~N.~C.; Babbush,~R.; Neven,~H. Decoding
  quantum errors with subspace expansions. \emph{Nat. Commun.} \textbf{2020},
  \emph{11}, 636\relax
\mciteBstWouldAddEndPuncttrue
\mciteSetBstMidEndSepPunct{\mcitedefaultmidpunct}
{\mcitedefaultendpunct}{\mcitedefaultseppunct}\relax
\EndOfBibitem
\bibitem[Nakanishi \latin{et~al.}(2019)Nakanishi, Mitarai, and
  Fujii]{nakanishi2019subspace}
Nakanishi,~K.~M.; Mitarai,~K.; Fujii,~K. Subspace-search variational quantum
  eigensolver for excited states. \emph{Phys. Rev. Research} \textbf{2019},
  \emph{1}, 033062\relax
\mciteBstWouldAddEndPuncttrue
\mciteSetBstMidEndSepPunct{\mcitedefaultmidpunct}
{\mcitedefaultendpunct}{\mcitedefaultseppunct}\relax
\EndOfBibitem
\bibitem[Parrish \latin{et~al.}(2019)Parrish, Hohenstein, McMahon, and
  Mart{\'\i}nez]{parrish2019quantum}
Parrish,~R.~M.; Hohenstein,~E.~G.; McMahon,~P.~L.; Mart{\'\i}nez,~T.~J. Quantum
  computation of electronic transitions using a variational quantum
  eigensolver. \emph{Phys. Rev. Lett.} \textbf{2019}, \emph{122}, 230401\relax
\mciteBstWouldAddEndPuncttrue
\mciteSetBstMidEndSepPunct{\mcitedefaultmidpunct}
{\mcitedefaultendpunct}{\mcitedefaultseppunct}\relax
\EndOfBibitem
\bibitem[Yalouz \latin{et~al.}(2021)Yalouz, Senjean, G{\"u}nther, Buda,
  O’Brien, and Visscher]{yalouz2021state}
Yalouz,~S.; Senjean,~B.; G{\"u}nther,~J.; Buda,~F.; O’Brien,~T.~E.;
  Visscher,~L. A state-averaged orbital-optimized hybrid quantum--classical
  algorithm for a democratic description of ground and excited states.
  \emph{Quantum Sci. Technol.} \textbf{2021}, \emph{6}, 024004\relax
\mciteBstWouldAddEndPuncttrue
\mciteSetBstMidEndSepPunct{\mcitedefaultmidpunct}
{\mcitedefaultendpunct}{\mcitedefaultseppunct}\relax
\EndOfBibitem
\bibitem[Fitzpatrick \latin{et~al.}(2024)Fitzpatrick, Nykanen, Talarico,
  Lunghi, Maniscalco, Garc{\'\i}a-P{\'e}rez, and Knecht]{fitzpatrick2024self}
Fitzpatrick,~A.; Nykanen,~A.; Talarico,~N.~W.; Lunghi,~A.; Maniscalco,~S.;
  Garc{\'\i}a-P{\'e}rez,~G.; Knecht,~S. Self-Consistent Field Approach for the
  Variational Quantum Eigensolver: Orbital Optimization Goes Adaptive. \emph{J.
  Phys. Chem. A} \textbf{2024}, \emph{128}, 2843--2856\relax
\mciteBstWouldAddEndPuncttrue
\mciteSetBstMidEndSepPunct{\mcitedefaultmidpunct}
{\mcitedefaultendpunct}{\mcitedefaultseppunct}\relax
\EndOfBibitem
\bibitem[Grimsley and Evangelista(2024)Grimsley, and
  Evangelista]{grimsley2024challenging}
Grimsley,~H.~R.; Evangelista,~F.~A. Challenging Excited States from Adaptive
  Quantum Eigensolvers: Subspace Expansions vs. State-Averaged Strategies.
  \emph{arXiv preprint arXiv:2409.11210} \textbf{2024}, \relax
\mciteBstWouldAddEndPunctfalse
\mciteSetBstMidEndSepPunct{\mcitedefaultmidpunct}
{}{\mcitedefaultseppunct}\relax
\EndOfBibitem
\bibitem[Higgott \latin{et~al.}(2019)Higgott, Wang, and
  Brierley]{higgott2019variational}
Higgott,~O.; Wang,~D.; Brierley,~S. Variational quantum computation of excited
  states. \emph{Quantum} \textbf{2019}, \emph{3}, 156\relax
\mciteBstWouldAddEndPuncttrue
\mciteSetBstMidEndSepPunct{\mcitedefaultmidpunct}
{\mcitedefaultendpunct}{\mcitedefaultseppunct}\relax
\EndOfBibitem
\bibitem[Tomasi \latin{et~al.}(2005)Tomasi, Mennucci, and
  Cammi]{tomasi2005quantum}
Tomasi,~J.; Mennucci,~B.; Cammi,~R. Quantum mechanical continuum solvation
  models. \emph{Chem. Rev.} \textbf{2005}, \emph{105}, 2999--3094\relax
\mciteBstWouldAddEndPuncttrue
\mciteSetBstMidEndSepPunct{\mcitedefaultmidpunct}
{\mcitedefaultendpunct}{\mcitedefaultseppunct}\relax
\EndOfBibitem
\bibitem[Singh and Kollman(1986)Singh, and Kollman]{singh1986combined}
Singh,~U.~C.; Kollman,~P.~A. A combined ab initio quantum mechanical and
  molecular mechanical method for carrying out simulations on complex molecular
  systems: {Applications} to the {CH3Cl} + Cl? exchange reaction and gas phase
  protonation of polyethers. \emph{J. Comput. Chem.} \textbf{1986}, \emph{7},
  718--730\relax
\mciteBstWouldAddEndPuncttrue
\mciteSetBstMidEndSepPunct{\mcitedefaultmidpunct}
{\mcitedefaultendpunct}{\mcitedefaultseppunct}\relax
\EndOfBibitem
\bibitem[Field \latin{et~al.}(1990)Field, Bash, and Karplus]{field1990combined}
Field,~M.~J.; Bash,~P.~A.; Karplus,~M. A combined quantum mechanical and
  molecular mechanical potential for molecular dynamics simulations. \emph{J.
  Comput. Chem.} \textbf{1990}, \emph{11}, 700--733\relax
\mciteBstWouldAddEndPuncttrue
\mciteSetBstMidEndSepPunct{\mcitedefaultmidpunct}
{\mcitedefaultendpunct}{\mcitedefaultseppunct}\relax
\EndOfBibitem
\bibitem[Olsen \latin{et~al.}(2010)Olsen, Aidas, and
  Kongsted]{olsen2010excited}
Olsen,~J.~M.; Aidas,~K.; Kongsted,~J. Excited states in solution through
  polarizable embedding. \emph{J. Chem. Theory Comput.} \textbf{2010},
  \emph{6}, 3721--3734\relax
\mciteBstWouldAddEndPuncttrue
\mciteSetBstMidEndSepPunct{\mcitedefaultmidpunct}
{\mcitedefaultendpunct}{\mcitedefaultseppunct}\relax
\EndOfBibitem
\bibitem[Olsen and Kongsted(2011)Olsen, and Kongsted]{olsen2011molecular}
Olsen,~J. M.~H.; Kongsted,~J. \emph{Advances in quantum chemistry}; Elsevier,
  2011; Vol.~61; pp 107--143\relax
\mciteBstWouldAddEndPuncttrue
\mciteSetBstMidEndSepPunct{\mcitedefaultmidpunct}
{\mcitedefaultendpunct}{\mcitedefaultseppunct}\relax
\EndOfBibitem
\bibitem[Thompson(1996)]{thompson1996qm}
Thompson,~M.~A. {QM/MMpol:} {A} Consistent Model for {Solute/Solvent}
  Polarization. Application to the Aqueous Solvation and Spectroscopy of
  Formaldehyde, Acetaldehyde, and Acetone. \emph{J. Phys. Chem.} \textbf{1996},
  \emph{100}, 14492--14507\relax
\mciteBstWouldAddEndPuncttrue
\mciteSetBstMidEndSepPunct{\mcitedefaultmidpunct}
{\mcitedefaultendpunct}{\mcitedefaultseppunct}\relax
\EndOfBibitem
\bibitem[Gordon \latin{et~al.}(2007)Gordon, Slipchenko, Li, and
  Jensen]{gordon2007effective}
Gordon,~M.~S.; Slipchenko,~L.; Li,~H.; Jensen,~J.~H. The Effective Fragment
  Potential: {A} General Method for Predicting Intermolecular Interactions.
  \emph{Annu Rep Comput Chem} \textbf{2007}, \emph{3}, 177--193\relax
\mciteBstWouldAddEndPuncttrue
\mciteSetBstMidEndSepPunct{\mcitedefaultmidpunct}
{\mcitedefaultendpunct}{\mcitedefaultseppunct}\relax
\EndOfBibitem
\bibitem[Loco \latin{et~al.}(2016)Loco, Polack, Caprasecca, Lagardère,
  Lipparini, Piquemal, and Mennucci]{loco2016qm}
Loco,~D.; Polack,~E.; Caprasecca,~S.; Lagardère,~L.; Lipparini,~F.;
  Piquemal,~J.-P.; Mennucci,~B. A {QM/MM} Approach Using the {AMOEBA}
  Polarizable Embedding: {From} Ground State Energies to Electronic
  Excitations. \emph{J. Chem. Theory Comput.} \textbf{2016}, \emph{12},
  3654--3661\relax
\mciteBstWouldAddEndPuncttrue
\mciteSetBstMidEndSepPunct{\mcitedefaultmidpunct}
{\mcitedefaultendpunct}{\mcitedefaultseppunct}\relax
\EndOfBibitem
\bibitem[Lipparini and Barone(2011)Lipparini, and Barone]{Lipparini2011}
Lipparini,~F.; Barone,~V. Polarizable Force Fields and Polarizable Continuum
  Model: {A} Fluctuating {Charges/PCM} Approach. 1. Theory and Implementation.
  \emph{J. Chem. Theory Comput.} \textbf{2011}, \emph{7}, 3711--3724\relax
\mciteBstWouldAddEndPuncttrue
\mciteSetBstMidEndSepPunct{\mcitedefaultmidpunct}
{\mcitedefaultendpunct}{\mcitedefaultseppunct}\relax
\EndOfBibitem
\bibitem[Lipparini \latin{et~al.}(2012)Lipparini, Cappelli, Scalmani, De~Mitri,
  and Barone]{lipparini2012analytical}
Lipparini,~F.; Cappelli,~C.; Scalmani,~G.; De~Mitri,~N.; Barone,~V. Analytical
  First and Second Derivatives for a Fully Polarizable {QM/Classical}
  {Hamiltonian}. \emph{J. Chem. Theory Comput.} \textbf{2012}, \emph{8},
  4270--4278\relax
\mciteBstWouldAddEndPuncttrue
\mciteSetBstMidEndSepPunct{\mcitedefaultmidpunct}
{\mcitedefaultendpunct}{\mcitedefaultseppunct}\relax
\EndOfBibitem
\bibitem[Lipparini \latin{et~al.}(2012)Lipparini, Cappelli, and
  Barone]{lipparini2012linear}
Lipparini,~F.; Cappelli,~C.; Barone,~V. Linear Response Theory and Electronic
  Transition Energies for a Fully Polarizable {QM/Classical} {Hamiltonian}.
  \emph{J. Chem. Theory Comput.} \textbf{2012}, \emph{8}, 4153--4165\relax
\mciteBstWouldAddEndPuncttrue
\mciteSetBstMidEndSepPunct{\mcitedefaultmidpunct}
{\mcitedefaultendpunct}{\mcitedefaultseppunct}\relax
\EndOfBibitem
\bibitem[Wesolowski and Weber(1997)Wesolowski, and Weber]{wesolowski1997kohn}
Wesolowski,~T.~A.; Weber,~J. Kohn-Sham equations with constrained electron
  density: {The} effect of various kinetic energy functional parametrizations
  on the ground-state molecular properties. \emph{Int. J. Quantum Chem.}
  \textbf{1997}, \emph{61}, 303--311\relax
\mciteBstWouldAddEndPuncttrue
\mciteSetBstMidEndSepPunct{\mcitedefaultmidpunct}
{\mcitedefaultendpunct}{\mcitedefaultseppunct}\relax
\EndOfBibitem
\bibitem[Manby \latin{et~al.}(2012)Manby, Stella, Goodpaster, and
  Miller]{manby2012simple}
Manby,~F.~R.; Stella,~M.; Goodpaster,~J.~D.; Miller,~T.~F. A Simple, Exact
  Density-Functional-Theory Embedding Scheme. \emph{J. Chem. Theory Comput.}
  \textbf{2012}, \emph{8}, 2564--2568\relax
\mciteBstWouldAddEndPuncttrue
\mciteSetBstMidEndSepPunct{\mcitedefaultmidpunct}
{\mcitedefaultendpunct}{\mcitedefaultseppunct}\relax
\EndOfBibitem
\bibitem[Myhre \latin{et~al.}(2014)Myhre, S{\'a}nchez~de Mer{\'a}s, and
  Koch]{myhre2014multi}
Myhre,~R.~H.; S{\'a}nchez~de Mer{\'a}s,~A.~M.; Koch,~H. Multi-level coupled
  cluster theory. \emph{J. Chem. Phys.} \textbf{2014}, \emph{141}\relax
\mciteBstWouldAddEndPuncttrue
\mciteSetBstMidEndSepPunct{\mcitedefaultmidpunct}
{\mcitedefaultendpunct}{\mcitedefaultseppunct}\relax
\EndOfBibitem
\bibitem[Castaldo \latin{et~al.}(2022)Castaldo, Jahangiri, Delgado, and
  Corni]{castaldo2022quantum}
Castaldo,~D.; Jahangiri,~S.; Delgado,~A.; Corni,~S. Quantum simulation of
  molecules in solution. \emph{J. Chem. Theory Comput.} \textbf{2022},
  \emph{18}, 7457--7469\relax
\mciteBstWouldAddEndPuncttrue
\mciteSetBstMidEndSepPunct{\mcitedefaultmidpunct}
{\mcitedefaultendpunct}{\mcitedefaultseppunct}\relax
\EndOfBibitem
\bibitem[Shee \latin{et~al.}(2023)Shee, Yeh, Hsiao, Yang, Lin, and
  Hsieh]{shee2023quantum}
Shee,~Y.; Yeh,~T.-L.; Hsiao,~J.-Y.; Yang,~A.; Lin,~Y.-C.; Hsieh,~M.-H. Quantum
  simulation of preferred tautomeric state prediction. \emph{npj Quantum Inf.}
  \textbf{2023}, \emph{9}, 102\relax
\mciteBstWouldAddEndPuncttrue
\mciteSetBstMidEndSepPunct{\mcitedefaultmidpunct}
{\mcitedefaultendpunct}{\mcitedefaultseppunct}\relax
\EndOfBibitem
\bibitem[Hohenstein \latin{et~al.}(2023)Hohenstein, Oumarou, Al-Saadon,
  Anselmetti, Scheurer, Gogolin, and Parrish]{hohenstein2023efficient}
Hohenstein,~E.~G.; Oumarou,~O.; Al-Saadon,~R.; Anselmetti,~G.-L.~R.;
  Scheurer,~M.; Gogolin,~C.; Parrish,~R.~M. Efficient quantum analytic nuclear
  gradients with double factorization. \emph{J. Chem. Phys.} \textbf{2023},
  \emph{158}\relax
\mciteBstWouldAddEndPuncttrue
\mciteSetBstMidEndSepPunct{\mcitedefaultmidpunct}
{\mcitedefaultendpunct}{\mcitedefaultseppunct}\relax
\EndOfBibitem
\bibitem[Kjellgren \latin{et~al.}(2024)Kjellgren, Reinholdt, Fitzpatrick,
  Talarico, Jensen, Sauer, Coriani, Knecht, and
  Kongsted]{kjellgren2024variational}
Kjellgren,~E.~R.; Reinholdt,~P.; Fitzpatrick,~A.; Talarico,~W.~N.;
  Jensen,~P.~W.; Sauer,~S.; Coriani,~S.; Knecht,~S.; Kongsted,~J. The
  variational quantum eigensolver self-consistent field method within a
  polarizable embedded framework. \emph{J. Chem. Phys.} \textbf{2024},
  \emph{160}\relax
\mciteBstWouldAddEndPuncttrue
\mciteSetBstMidEndSepPunct{\mcitedefaultmidpunct}
{\mcitedefaultendpunct}{\mcitedefaultseppunct}\relax
\EndOfBibitem
\bibitem[Nagy \latin{et~al.}(2024)Nagy, Reinholdt, Jensen, Kjellgren, Ziems,
  Fitzpatrick, Knecht, Kongsted, Coriani, and Sauer]{nagy2024electric}
Nagy,~D.; Reinholdt,~P.; Jensen,~P.~W.; Kjellgren,~E.~R.; Ziems,~K.~M.;
  Fitzpatrick,~A.; Knecht,~S.; Kongsted,~J.; Coriani,~S.; Sauer,~S.~P. Electric
  Field Gradient Calculations for Ice VIII and IX Using Polarizable Embedding:
  A Comparative Study on Classical Computers and Quantum Simulators. \emph{J.
  Phys. Chem. A} \textbf{2024}, \emph{128}, 6305--6315\relax
\mciteBstWouldAddEndPuncttrue
\mciteSetBstMidEndSepPunct{\mcitedefaultmidpunct}
{\mcitedefaultendpunct}{\mcitedefaultseppunct}\relax
\EndOfBibitem
\bibitem[Rossmannek \latin{et~al.}(2023)Rossmannek, Pavosevic, Rubio, and
  Tavernelli]{rossmannek2023quantum}
Rossmannek,~M.; Pavosevic,~F.; Rubio,~A.; Tavernelli,~I. Quantum embedding
  method for the simulation of strongly correlated systems on quantum
  computers. \emph{J. Phys. Chem. Lett.} \textbf{2023}, \emph{14},
  3491--3497\relax
\mciteBstWouldAddEndPuncttrue
\mciteSetBstMidEndSepPunct{\mcitedefaultmidpunct}
{\mcitedefaultendpunct}{\mcitedefaultseppunct}\relax
\EndOfBibitem
\bibitem[Weisburn \latin{et~al.}(2024)Weisburn, Cho, Bensberg, Meitei, Reiher,
  and Van~Voorhis]{weisburn2024multiscale}
Weisburn,~L.~P.; Cho,~M.; Bensberg,~M.; Meitei,~O.~R.; Reiher,~M.;
  Van~Voorhis,~T. Multiscale Embedding for Quantum Computing. \emph{arXiv
  preprint arXiv:2409.06813} \textbf{2024}, \relax
\mciteBstWouldAddEndPunctfalse
\mciteSetBstMidEndSepPunct{\mcitedefaultmidpunct}
{}{\mcitedefaultseppunct}\relax
\EndOfBibitem
\bibitem[Ettenhuber \latin{et~al.}(2024)Ettenhuber, Hansen, Shaik, Rasmussen,
  Poier, Madsen, Majland, Jensen, Olsen, and Zinner]{ettenhuber2024calculating}
Ettenhuber,~P.; Hansen,~M.~B.; Shaik,~I.; Rasmussen,~S.~E.; Poier,~P.~P.;
  Madsen,~N.~K.; Majland,~M.; Jensen,~F.; Olsen,~L.; Zinner,~N.~T. Calculating
  the energy profile of an enzymatic reaction on a quantum computer.
  \emph{arXiv preprint arXiv:2408.11091} \textbf{2024}, \relax
\mciteBstWouldAddEndPunctfalse
\mciteSetBstMidEndSepPunct{\mcitedefaultmidpunct}
{}{\mcitedefaultseppunct}\relax
\EndOfBibitem
\bibitem[Battaglia \latin{et~al.}(2024)Battaglia, Rossmannek, Rybkin,
  Tavernelli, and Hutter]{battaglia2024general}
Battaglia,~S.; Rossmannek,~M.; Rybkin,~V.~V.; Tavernelli,~I.; Hutter,~J. A
  general framework for active space embedding methods: applications in quantum
  computing. \emph{arXiv preprint arXiv:2404.18737} \textbf{2024}, \relax
\mciteBstWouldAddEndPunctfalse
\mciteSetBstMidEndSepPunct{\mcitedefaultmidpunct}
{}{\mcitedefaultseppunct}\relax
\EndOfBibitem
\bibitem[Paldus \latin{et~al.}(1977)Paldus, Adams, and Čížek]{Paldus1977}
Paldus,~J.; Adams,~B.~G.; Čížek,~J. Application of graphical methods of spin
  algebras to limited {CI} approaches. I. Closed shell case. \emph{Int. J.
  Quantum Chem.} \textbf{1977}, \emph{11}, 813–848\relax
\mciteBstWouldAddEndPuncttrue
\mciteSetBstMidEndSepPunct{\mcitedefaultmidpunct}
{\mcitedefaultendpunct}{\mcitedefaultseppunct}\relax
\EndOfBibitem
\bibitem[Piecuch and Paldus(1989)Piecuch, and Paldus]{Piecuch1989}
Piecuch,~P.; Paldus,~J. Orthogonally spin‐adapted coupled‐cluster equations
  involving singly and doubly excited clusters. Comparison of different
  procedures for spin‐adaptation. \emph{Int. J. Quantum Chem.} \textbf{1989},
  \emph{36}, 429–453\relax
\mciteBstWouldAddEndPuncttrue
\mciteSetBstMidEndSepPunct{\mcitedefaultmidpunct}
{\mcitedefaultendpunct}{\mcitedefaultseppunct}\relax
\EndOfBibitem
\bibitem[Packer \latin{et~al.}(1996)Packer, Dalskov, Enevoldsen, Jensen, and
  Oddershede]{Packer1996}
Packer,~M.~J.; Dalskov,~E.~K.; Enevoldsen,~T.; Jensen,~H. J.~A.; Oddershede,~J.
  A new implementation of the second-order polarization propagator
  approximation ({SOPPA}): The excitation spectra of benzene and naphthalene.
  \emph{J. Chem. Phys.} \textbf{1996}, \emph{105}, 5886--5900\relax
\mciteBstWouldAddEndPuncttrue
\mciteSetBstMidEndSepPunct{\mcitedefaultmidpunct}
{\mcitedefaultendpunct}{\mcitedefaultseppunct}\relax
\EndOfBibitem
\bibitem[Taube and Bartlett(2006)Taube, and Bartlett]{taube2006new}
Taube,~A.~G.; Bartlett,~R.~J. New perspectives on unitary coupled-cluster
  theory. \emph{Int. J. Quantum Chem.} \textbf{2006}, \emph{106},
  3393--3401\relax
\mciteBstWouldAddEndPuncttrue
\mciteSetBstMidEndSepPunct{\mcitedefaultmidpunct}
{\mcitedefaultendpunct}{\mcitedefaultseppunct}\relax
\EndOfBibitem
\bibitem[Davidson(1975)]{DAVIDSON197587}
Davidson,~E.~R. The iterative calculation of a few of the lowest eigenvalues
  and corresponding eigenvectors of large real-symmetric matrices. \emph{J.
  Comput. Phys.} \textbf{1975}, \emph{17}, 87--94\relax
\mciteBstWouldAddEndPuncttrue
\mciteSetBstMidEndSepPunct{\mcitedefaultmidpunct}
{\mcitedefaultendpunct}{\mcitedefaultseppunct}\relax
\EndOfBibitem
\bibitem[Olsen \latin{et~al.}(1988)Olsen, Jensen, and
  J{\o}rgensen]{olsen1988solution}
Olsen,~J.; Jensen,~H. J.~A.; J{\o}rgensen,~P. Solution of the large matrix
  equations which occur in response theory. \emph{J. Comput. Phys.}
  \textbf{1988}, \emph{74}, 265--282\relax
\mciteBstWouldAddEndPuncttrue
\mciteSetBstMidEndSepPunct{\mcitedefaultmidpunct}
{\mcitedefaultendpunct}{\mcitedefaultseppunct}\relax
\EndOfBibitem
\bibitem[Kim and Krylov(2023)Kim, and Krylov]{kim2023two}
Kim,~Y.; Krylov,~A.~I. Two Algorithms for Excited-State Quantum Solvers: Theory
  and Application to EOM-UCCSD. \emph{J. Phys. Chem. A} \textbf{2023},
  \emph{127}, 6552--6566\relax
\mciteBstWouldAddEndPuncttrue
\mciteSetBstMidEndSepPunct{\mcitedefaultmidpunct}
{\mcitedefaultendpunct}{\mcitedefaultseppunct}\relax
\EndOfBibitem
\bibitem[Mottonen \latin{et~al.}(2004)Mottonen, Vartiainen, Bergholm, and
  Salomaa]{mottonen2004transformation}
Mottonen,~M.; Vartiainen,~J.~J.; Bergholm,~V.; Salomaa,~M.~M. Transformation of
  quantum states using uniformly controlled rotations. \emph{arXiv preprint
  quant-ph/0407010} \textbf{2004}, \relax
\mciteBstWouldAddEndPunctfalse
\mciteSetBstMidEndSepPunct{\mcitedefaultmidpunct}
{}{\mcitedefaultseppunct}\relax
\EndOfBibitem
\bibitem[Bergholm \latin{et~al.}(2022)Bergholm, Izaac, Schuld, Gogolin, Ahmed,
  Ajith, Alam, Alonso-Linaje, AkashNarayanan, Asadi, Arrazola, Azad, Banning,
  Blank, Bromley, Cordier, Ceroni, Delgado, Matteo, Dusko, Garg, Guala, Hayes,
  Hill, Ijaz, Isacsson, Ittah, Jahangiri, Jain, Jiang, Khandelwal, Kottmann,
  Lang, Lee, Loke, Lowe, McKiernan, Meyer, Montañez-Barrera, Moyard, Niu,
  O'Riordan, Oud, Panigrahi, Park, Polatajko, Quesada, Roberts, Sá, Schoch,
  Shi, Shu, Sim, Singh, Strandberg, Soni, Száva, Thabet, Vargas-Hernández,
  Vincent, Vitucci, Weber, Wierichs, Wiersema, Willmann, Wong, Zhang, and
  Killoran]{pennylane}
Bergholm,~V.; Izaac,~J.; Schuld,~M.; Gogolin,~C.; Ahmed,~S.; Ajith,~V.;
  Alam,~M.~S.; Alonso-Linaje,~G.; AkashNarayanan,~B.; Asadi,~A. \latin{et~al.}
  PennyLane: Automatic differentiation of hybrid quantum-classical
  computations. \emph{arXiv:1811.04968 (quant-ph)} \textbf{2022}, \relax
\mciteBstWouldAddEndPunctfalse
\mciteSetBstMidEndSepPunct{\mcitedefaultmidpunct}
{}{\mcitedefaultseppunct}\relax
\EndOfBibitem
\bibitem[Stone(2013)]{stone2013theory}
Stone,~A. \emph{{The Theory of Intermolecular Forces}}; Oxford University
  Press, 2013\relax
\mciteBstWouldAddEndPuncttrue
\mciteSetBstMidEndSepPunct{\mcitedefaultmidpunct}
{\mcitedefaultendpunct}{\mcitedefaultseppunct}\relax
\EndOfBibitem
\bibitem[List \latin{et~al.}(2016)List, Jensen, and Kongsted]{list2016local}
List,~N.~H.; Jensen,~H. J.~A.; Kongsted,~J. Local electric fields and molecular
  properties in heterogeneous environments through polarizable embedding.
  \emph{Phys. Chem. Chem. Phys.} \textbf{2016}, \emph{18}, 10070--10080\relax
\mciteBstWouldAddEndPuncttrue
\mciteSetBstMidEndSepPunct{\mcitedefaultmidpunct}
{\mcitedefaultendpunct}{\mcitedefaultseppunct}\relax
\EndOfBibitem
\bibitem[Ziems \latin{et~al.}(2024)Ziems, Kjellgren, Sauer, Kongsted, and
  Coriani]{ziems2024understanding}
Ziems,~K.~M.; Kjellgren,~E.~R.; Sauer,~S.; Kongsted,~J.; Coriani,~S.
  Understanding and mitigating noise in molecular quantum linear response for
  spectroscopic properties on quantum computers. \emph{arXiv preprint
  arXiv:2408.09308} \textbf{2024}, \relax
\mciteBstWouldAddEndPunctfalse
\mciteSetBstMidEndSepPunct{\mcitedefaultmidpunct}
{}{\mcitedefaultseppunct}\relax
\EndOfBibitem
\bibitem[Maciejewski \latin{et~al.}(2020)Maciejewski, Zimbor{\'a}s, and
  Oszmaniec]{maciejewski2020mitigation}
Maciejewski,~F.~B.; Zimbor{\'a}s,~Z.; Oszmaniec,~M. Mitigation of readout noise
  in near-term quantum devices by classical post-processing based on detector
  tomography. \emph{Quantum} \textbf{2020}, \emph{4}, 257\relax
\mciteBstWouldAddEndPuncttrue
\mciteSetBstMidEndSepPunct{\mcitedefaultmidpunct}
{\mcitedefaultendpunct}{\mcitedefaultseppunct}\relax
\EndOfBibitem
\bibitem[Motta \latin{et~al.}(2023)Motta, Jones, Rice, Gujarati, Sakuma,
  Liepuoniute, Garcia, and Ohnishi]{motta2023quantum}
Motta,~M.; Jones,~G.~O.; Rice,~J.~E.; Gujarati,~T.~P.; Sakuma,~R.;
  Liepuoniute,~I.; Garcia,~J.~M.; Ohnishi,~Y.-y. Quantum chemistry simulation
  of ground-and excited-state properties of the sulfonium cation on a
  superconducting quantum processor. \emph{Chem. Sci.} \textbf{2023},
  \emph{14}, 2915--2927\relax
\mciteBstWouldAddEndPuncttrue
\mciteSetBstMidEndSepPunct{\mcitedefaultmidpunct}
{\mcitedefaultendpunct}{\mcitedefaultseppunct}\relax
\EndOfBibitem
\bibitem[Javadi-Abhari \latin{et~al.}(2024)Javadi-Abhari, Treinish, Krsulich,
  Wood, Lishman, Gacon, Martiel, Nation, Bishop, Cross, Johnson, and
  Gambetta]{qiskit2024}
Javadi-Abhari,~A.; Treinish,~M.; Krsulich,~K.; Wood,~C.~J.; Lishman,~J.;
  Gacon,~J.; Martiel,~S.; Nation,~P.~D.; Bishop,~L.~S.; Cross,~A.~W.
  \latin{et~al.}  Quantum computing with {Q}iskit. 2024\relax
\mciteBstWouldAddEndPuncttrue
\mciteSetBstMidEndSepPunct{\mcitedefaultmidpunct}
{\mcitedefaultendpunct}{\mcitedefaultseppunct}\relax
\EndOfBibitem
\bibitem[Sun \latin{et~al.}(2020)Sun, Zhang, Banerjee, Bao, Barbry, Blunt,
  Bogdanov, Booth, Chen, Cui, \latin{et~al.} others]{pyscf}
Sun,~Q.; Zhang,~X.; Banerjee,~S.; Bao,~P.; Barbry,~M.; Blunt,~N.~S.;
  Bogdanov,~N.~A.; Booth,~G.~H.; Chen,~J.; Cui,~Z.-H. \latin{et~al.}  Recent
  developments in the PySCF program package. \emph{J. Chem. Phys.}
  \textbf{2020}, \emph{153}\relax
\mciteBstWouldAddEndPuncttrue
\mciteSetBstMidEndSepPunct{\mcitedefaultmidpunct}
{\mcitedefaultendpunct}{\mcitedefaultseppunct}\relax
\EndOfBibitem
\bibitem[Scheurer \latin{et~al.}(2019)Scheurer, Reinholdt, Kjellgren,
  Haugaard~Olsen, Dreuw, and Kongsted]{scheurer2019cppe}
Scheurer,~M.; Reinholdt,~P.; Kjellgren,~E.~R.; Haugaard~Olsen,~J.~M.;
  Dreuw,~A.; Kongsted,~J. CPPE: An open-source C++ and Python library for
  polarizable embedding. \emph{J. Chem. Theory Comput.} \textbf{2019},
  \emph{15}, 6154--6163\relax
\mciteBstWouldAddEndPuncttrue
\mciteSetBstMidEndSepPunct{\mcitedefaultmidpunct}
{\mcitedefaultendpunct}{\mcitedefaultseppunct}\relax
\EndOfBibitem
\bibitem[McClean \latin{et~al.}(2020)McClean, Rubin, Sung, Kivlichan,
  Bonet-Monroig, Cao, Dai, Fried, Gidney, Gimby, \latin{et~al.}
  others]{mcclean2020openfermion}
McClean,~J.~R.; Rubin,~N.~C.; Sung,~K.~J.; Kivlichan,~I.~D.; Bonet-Monroig,~X.;
  Cao,~Y.; Dai,~C.; Fried,~E.~S.; Gidney,~C.; Gimby,~B. \latin{et~al.}
  OpenFermion: the electronic structure package for quantum computers.
  \emph{Quantum Sci. Technol.} \textbf{2020}, \emph{5}, 034014\relax
\mciteBstWouldAddEndPuncttrue
\mciteSetBstMidEndSepPunct{\mcitedefaultmidpunct}
{\mcitedefaultendpunct}{\mcitedefaultseppunct}\relax
\EndOfBibitem
\bibitem[Kraft(1988)]{slsqp}
Kraft,~D. A software package for sequential quadratic programming.
  \emph{Forschungsbericht- Deutsche Forschungs- und Versuchsanstalt fur Luft-
  und Raumfahrt} \textbf{1988}, \relax
\mciteBstWouldAddEndPunctfalse
\mciteSetBstMidEndSepPunct{\mcitedefaultmidpunct}
{}{\mcitedefaultseppunct}\relax
\EndOfBibitem
\bibitem[Virtanen \latin{et~al.}(2020)Virtanen, Gommers, Oliphant, Haberland,
  Reddy, Cournapeau, Burovski, Peterson, Weckesser, Bright, van~der Walt,
  Brett, Wilson, Millman, Mayorov, Nelson, Jones, Kern, Larson, Carey, Polat,
  Feng, Moore, VanderPlas, Laxalde, Perktold, Cimrman, Henriksen, Quintero,
  Harris, Archibald, Ribeiro, Pedregosa, van Mulbregt, Vijaykumar, Bardelli,
  Rothberg, Hilboll, Kloeckner, Scopatz, Lee, Rokem, Woods, Fulton, Masson,
  H\"{a}ggstr\"{o}m, Fitzgerald, Nicholson, Hagen, Pasechnik, Olivetti, Martin,
  Wieser, Silva, Lenders, Wilhelm, Young, Price, Ingold, Allen, Lee, Audren,
  Probst, Dietrich, Silterra, Webber, Slavi\v{c}, Nothman, Buchner, Kulick,
  Sch\"{o}nberger, de~Miranda~Cardoso, Reimer, Harrington, Rodr\'{\i}guez,
  Nunez-Iglesias, Kuczynski, Tritz, Thoma, Newville, K\"{u}mmerer, Bolingbroke,
  Tartre, Pak, Smith, Nowaczyk, Shebanov, Pavlyk, Brodtkorb, Lee, McGibbon,
  Feldbauer, Lewis, Tygier, Sievert, Vigna, Peterson, More, Pudlik, Oshima,
  Pingel, Robitaille, Spura, Jones, Cera, Leslie, Zito, Krauss, Upadhyay,
  Halchenko, V\'{a}zquez-Baeza, and {SciPy 1.0
  Contributors.}]{virtanen2020scipy}
Virtanen,~P.; Gommers,~R.; Oliphant,~T.~E.; Haberland,~M.; Reddy,~T.;
  Cournapeau,~D.; Burovski,~E.; Peterson,~P.; Weckesser,~W.; Bright,~J.
  \latin{et~al.}  SciPy 1.0: fundamental algorithms for scientific computing in
  Python. \emph{Nat. Methods} \textbf{2020}, \emph{17}, 261--272\relax
\mciteBstWouldAddEndPuncttrue
\mciteSetBstMidEndSepPunct{\mcitedefaultmidpunct}
{\mcitedefaultendpunct}{\mcitedefaultseppunct}\relax
\EndOfBibitem
\bibitem[Aidas \latin{et~al.}(2014)Aidas, Angeli, Bak, Bakken, Bast, Boman,
  Christiansen, Cimiraglia, Coriani, Dahle, Dalskov, Ekstr\"{o}m, Enevoldsen,
  Eriksen, Ettenhuber, Fern\'{a}ndez, Ferrighi, Fliegl, Frediani, Hald,
  Halkier, H\"{a}ttig, Heiberg, Helgaker, Hennum, Hettema, Hjerten\ae{}s,
  H\o{}st, H\o{}yvik, Iozzi, Jans\'{i}k, Jensen, Jonsson, J\o{}rgensen,
  Kauczor, Kirpekar, Kj\ae{}rgaard, Klopper, Knecht, Kobayashi, Koch, Kongsted,
  Krapp, Kristensen, Ligabue, Lutn\ae{}s, Melo, Mikkelsen, Myhre, Neiss,
  Nielsen, Norman, Olsen, Olsen, Osted, Packer, Pawlowski, Pedersen, Provasi,
  Reine, Rinkevicius, Ruden, Ruud, Rybkin, Sa\l{}ek, Samson, de~Mer\'{a}s,
  Saue, Sauer, Schimmelpfennig, Sneskov, Steindal, Sylvester-Hvid, Taylor,
  Teale, Tellgren, Tew, Thorvaldsen, Th\o{}gersen, Vahtras, Watson, Wilson,
  Ziolkowski, and \AA{}gren]{daltonpaper}
Aidas,~K.; Angeli,~C.; Bak,~K.~L.; Bakken,~V.; Bast,~R.; Boman,~L.;
  Christiansen,~O.; Cimiraglia,~R.; Coriani,~S.; Dahle,~P. \latin{et~al.}  {The
  Dalton quantum chemistry program system}. \emph{WIREs Comput.~Mol.~Sci.}
  \textbf{2014}, \emph{4}, 269--284\relax
\mciteBstWouldAddEndPuncttrue
\mciteSetBstMidEndSepPunct{\mcitedefaultmidpunct}
{\mcitedefaultendpunct}{\mcitedefaultseppunct}\relax
\EndOfBibitem
\bibitem[Reinholdt \latin{et~al.}(2021)Reinholdt, Kongsted, and
  Lipparini]{reinholdt2021fast}
Reinholdt,~P.; Kongsted,~J.; Lipparini,~F. Fast approximate but accurate QM/MM
  interactions for polarizable embedding. \emph{J. Chem. Theory Comput.}
  \textbf{2021}, \emph{18}, 344--356\relax
\mciteBstWouldAddEndPuncttrue
\mciteSetBstMidEndSepPunct{\mcitedefaultmidpunct}
{\mcitedefaultendpunct}{\mcitedefaultseppunct}\relax
\EndOfBibitem
\bibitem[Olsen and Reinholdt(2021)Olsen, and Reinholdt]{pyframe}
Olsen,~J. M.~H.; Reinholdt,~P. {PyFraME: Python framework for Fragment-based
  Multiscale Embedding}. 2021;
  \url{https://doi.org/10.5281/zenodo.4899311}\relax
\mciteBstWouldAddEndPuncttrue
\mciteSetBstMidEndSepPunct{\mcitedefaultmidpunct}
{\mcitedefaultendpunct}{\mcitedefaultseppunct}\relax
\EndOfBibitem
\bibitem[Gagliardi \latin{et~al.}(2004)Gagliardi, Lindh, and
  Karlstr{\"o}m]{gagliardi2004local}
Gagliardi,~L.; Lindh,~R.; Karlstr{\"o}m,~G. Local properties of quantum
  chemical systems: The LoProp approach. \emph{J. Chem. Phys.} \textbf{2004},
  \emph{121}, 4494--4500\relax
\mciteBstWouldAddEndPuncttrue
\mciteSetBstMidEndSepPunct{\mcitedefaultmidpunct}
{\mcitedefaultendpunct}{\mcitedefaultseppunct}\relax
\EndOfBibitem
\bibitem[Yanai \latin{et~al.}(2004)Yanai, Tew, and Handy]{yanai2004new}
Yanai,~T.; Tew,~D.~P.; Handy,~N.~C. A new hybrid exchange--correlation
  functional using the Coulomb-attenuating method (CAM-B3LYP). \emph{Chem.
  Phys. Lett.} \textbf{2004}, \emph{393}, 51--57\relax
\mciteBstWouldAddEndPuncttrue
\mciteSetBstMidEndSepPunct{\mcitedefaultmidpunct}
{\mcitedefaultendpunct}{\mcitedefaultseppunct}\relax
\EndOfBibitem
\bibitem[Van~den Heuvel \latin{et~al.}(2023)Van~den Heuvel, Reinholdt, and
  Kongsted]{van2023embedding}
Van~den Heuvel,~W.; Reinholdt,~P.; Kongsted,~J. Embedding Beyond
  Electrostatics: The Extended Polarizable Density Embedding Model. \emph{J.
  Phys. Chem. B} \textbf{2023}, \emph{127}, 3248--3256\relax
\mciteBstWouldAddEndPuncttrue
\mciteSetBstMidEndSepPunct{\mcitedefaultmidpunct}
{\mcitedefaultendpunct}{\mcitedefaultseppunct}\relax
\EndOfBibitem
\bibitem[Abraham \latin{et~al.}(2015)Abraham, Murtola, Schulz, P{\'a}ll, Smith,
  Hess, and Lindahl]{abraham2015gromacs}
Abraham,~M.~J.; Murtola,~T.; Schulz,~R.; P{\'a}ll,~S.; Smith,~J.~C.; Hess,~B.;
  Lindahl,~E. GROMACS: High performance molecular simulations through
  multi-level parallelism from laptops to supercomputers. \emph{SoftwareX}
  \textbf{2015}, \emph{1}, 19--25\relax
\mciteBstWouldAddEndPuncttrue
\mciteSetBstMidEndSepPunct{\mcitedefaultmidpunct}
{\mcitedefaultendpunct}{\mcitedefaultseppunct}\relax
\EndOfBibitem
\bibitem[Darden \latin{et~al.}(1993)Darden, York, and
  Pedersen]{darden1993particle}
Darden,~T.; York,~D.; Pedersen,~L. Particle mesh Ewald: An N log (N) method for
  Ewald sums in large systems. \emph{J. Chem. Phys.} \textbf{1993}, \emph{98},
  10089--10092\relax
\mciteBstWouldAddEndPuncttrue
\mciteSetBstMidEndSepPunct{\mcitedefaultmidpunct}
{\mcitedefaultendpunct}{\mcitedefaultseppunct}\relax
\EndOfBibitem
\bibitem[Hess \latin{et~al.}(1997)Hess, Bekker, Berendsen, and
  Fraaije]{hess1997lincs}
Hess,~B.; Bekker,~H.; Berendsen,~H.~J.; Fraaije,~J.~G. LINCS: a linear
  constraint solver for molecular simulations. \emph{J. Comput. Chem.}
  \textbf{1997}, \emph{18}, 1463--1472\relax
\mciteBstWouldAddEndPuncttrue
\mciteSetBstMidEndSepPunct{\mcitedefaultmidpunct}
{\mcitedefaultendpunct}{\mcitedefaultseppunct}\relax
\EndOfBibitem
\bibitem[Bussi \latin{et~al.}(2007)Bussi, Donadio, and
  Parrinello]{bussi2007canonical}
Bussi,~G.; Donadio,~D.; Parrinello,~M. Canonical sampling through velocity
  rescaling. \emph{J. Chem. Phys.} \textbf{2007}, \emph{126}\relax
\mciteBstWouldAddEndPuncttrue
\mciteSetBstMidEndSepPunct{\mcitedefaultmidpunct}
{\mcitedefaultendpunct}{\mcitedefaultseppunct}\relax
\EndOfBibitem
\bibitem[Bernetti and Bussi(2020)Bernetti, and Bussi]{crescale}
Bernetti,~M.; Bussi,~G. {Pressure control using stochastic cell rescaling}.
  \emph{J. Chem. Phys.} \textbf{2020}, \emph{153}, 114107\relax
\mciteBstWouldAddEndPuncttrue
\mciteSetBstMidEndSepPunct{\mcitedefaultmidpunct}
{\mcitedefaultendpunct}{\mcitedefaultseppunct}\relax
\EndOfBibitem
\bibitem[Nanda and Krylov(2018)Nanda, and Krylov]{nanda2018effect}
Nanda,~K.~D.; Krylov,~A.~I. The effect of polarizable environment on two-photon
  absorption cross sections characterized by the equation-of-motion
  coupled-cluster singles and doubles method combined with the effective
  fragment potential approach. \emph{J. Chem. Phys.} \textbf{2018},
  \emph{149}\relax
\mciteBstWouldAddEndPuncttrue
\mciteSetBstMidEndSepPunct{\mcitedefaultmidpunct}
{\mcitedefaultendpunct}{\mcitedefaultseppunct}\relax
\EndOfBibitem
\end{mcitethebibliography}
\end{document}